\documentclass[reviewcopy]{elsart}

\usepackage{natbib}
\usepackage{graphicx}
\usepackage{amssymb}
\usepackage{amsmath}
\usepackage{latexsym}
\usepackage{longtable}

\newcommand{\ti}{$\Gamma$}
\newcommand{\tiu}{J m$^{-2}$ s$^{-0.5}$ K$^{-1}$}
\newcommand{\tb}{$\overline{\theta}$}

\newcommand{\TI}{thermal inertia}
\newcommand{\TIv}{thermal inertia value}
\newcommand{\TIs}{thermal inertia values}
\newcommand{\TIvs}{thermal inertia values}
\newcommand{\De}{$D$}
\newcommand{\lp}{$\lambda_p$}
\newcommand{\bp}{$\beta_p$}

\newcommand{\de}{$^\circ$}
\newcommand{\CS}{$\overline \chi^2$}
\newcommand{\co}{$\gamma_C$}
\newcommand{\cd}{$\rho_C$}
\newcommand{\OCA}{Observatoire de la C\^ote d'Azur}
\newcommand{\MMP}{Top: AMLI shape and spin vector solution 1, bottom: AMLI solution 2}
\newcommand{\SOM}{Supplementary On--line Material}
\newcommand{\hl}{}

\begin{document}

\begin{frontmatter}



\title{Thermal inertia of main belt asteroids smaller than 100 km from IRAS data}


\author{Marco Delbo'\corauthref{cor}\thanksref{md}}
\ead{delbo@obs-nice.fr} \corauth[cor]{Corresponding author.}
and
\author{Paolo Tanga}
\journal{Planetary and Space Science}

\address{Laboratoire Cassiop\'ee, Observatoire de la C\^ote d'Azur\\
BP 4229, 06304 Nice cedex 04, France.}

\thanks[md]{Supported by the European Space Agency (ESA). Also at INAF, Astronomical Observatory of Torino, Italy.}

\begin{abstract}
Recent works have shown that the \TI~of km-sized near-Earth asteroids (NEAs) 
is more than two orders of magnitude higher than that of main belt asteroids 
(MBAs) with sizes (diameters) between 200 and 1,000 km.
This confirms the idea that large MBAs, over hundreds millions of years,
have developed a fine and thick thermally insulating regolith layer, responsible for
the low values of their \TI, whereas km-sized asteroids, having collisional lifetimes
of only some millions years, have less regolith, and consequently a larger surface \TI.

Because it is believed that regolith on asteroids forms as a result of impact processes,
a better knowledge of asteroid \TI~and its correlation with size, taxonomic type,
and density can be used as an important constraint for modeling of impact processes on asteroids.
However, our knowledge of asteroids' \TIvs~is still based on few data points with NEAs 
covering the size range 0.1--20 km and MBAs that $>$100 km.

Here, we use IRAS infrared measurements to estimate the \TIs~of MBAs with 
diameters $<$100 km and known shapes and spin vector:
{\hl filling an important size gap between the largest MBAs and the km-sized NEAs.} 
An update to the inverse correlation between \TI~and diameter is presented.
{\hl For some asteroids thermophysical modelling allowed
us to discriminate between the two still possible spin vector solutions 
derived from optical lightcurve inversion. This is important for (720) Bohlinia: our
preferred solution was predicted to be the correct one 
by Vokrouhlick\'y et al. (2003, Nature 425, 147) just on theoretical grounds.}

\end{abstract}

\begin{keyword}
Asteroids \sep Near-Earth Objects \sep Infrared observations
\end{keyword}
\end{frontmatter}

%
%
\section{Introduction}

Thermal inertia is a measure of the resistance of a material to temperature change. It is defined by $\Gamma = \sqrt{\rho \kappa c}$, where $\kappa$ is the thermal conductivity, $\rho$ the density and $c$ the specific heat.
\ti~is a key parameter that controls the temperature distribution over the surface of an asteroid. In the limit of zero thermal inertia the surface of an asteroid is in instantaneous equilibrium with the solar radiation and displays a prominent temperature maximum at the sub-solar point. In the realistic case of a rotating asteroid with finite thermal inertia the temperature distribution becomes more smoothed out in longitude with the afternoon hemisphere hotter than the morning one \citep[see e.g.][and references therein]{Delbo02, Delbo04, Migo07}.

{\hl Acquisition of temperature data (e.g. from thermal infrared observations at different wavelengths around the body's heat emission peak) over a portion of the diurnal warming/cooling cycle can be used to derive the thermal inertia of planetary surfaces by fitting a temperature curve calculated by means of a thermal model to the observed data. Asteroids surface temperatures depend also on the bodies' shapes, inclination of their spin axis and rotation rates. 
For those objects for which this information is available the so-called thermophysical models (TMPs) can be used to calculate infrared fluxes as function of the asteroid's albedo, thermal inertia and macroscopic roughness. Those parameters are adjusted until best fit to the data
is obtained \citep[see][and \S \ref{S_method} for details]{Migo07}}

Knowledge of the thermal inertia of asteroid surfaces is important for several reasons:
\begin{enumerate}
\item  \TI~is a sensitive indicator for the presence or absence of {\hl thermally insulating} loose material on the surface such as regolith or dust \citep[see e.g.][]{Chris03}. The value of \ti~depends on {\hl regolith depth, degree of induration and particle size}, rock abundance, and exposure of solid rocks and boulders within the top few centimeters of the subsurface {\hl(i.e. a few thermal skin depths)}. Typical values of \ti~in (S.I. units \tiu) are 30 for fine dusts, 50 for the lunar regolith, 400 for coarse sands {\hl (note that a \TI~of 400 for coarse sand assumes the presence of some atmosphere, even if as thin as the Martian one)}, and 2500 for bare solid rocks
    \citep[][see also http://tes.asu.edu/TESworkshop/Mellon.pdf]{Mellon00, Spencer89, 1986Icar...66..117J}. Information about \TI~is therefore of great importance in the design of instrumentation for lander missions to asteroids such as the Marco Polo of the European Space Agency, because it allows one to have information about the soil and sub--soil temperatures and the make up of asteroid surfaces.

\item The presence or absence and thickness of the regolith on km--sized bodies can give hints about the internal structure of asteroids: recent work by \citet{2007P&SS...55...70M} showed that small asteroids (with sizes $\sim$1 km) can capture collisional debris and build up regolith if their tensile strength is not high;

\item Thermal inertia affects the strength of the Yarkovsky effect \citep[see][and references therein]{Bottke06} which is responsible for the gradual drifting of the orbits of km-sized asteroids and is thought to play an important role in the delivery of near-Earth asteroids (NEAs) from the main belt \citep{Morby03}, and in the dynamical spreading of asteroid families \citep[see][]{Bottke06}.

\item Understanding asteroid \TI~is important to estimate and reduce systematic errors on sizes and albedos of asteroids, when the these are determined by means of simple thermal models such as the Standard Thermal Model \citep[STM;][]{Lebo89} neglecting the effect of the rotation of these bodies and their thermal inertia \citep{Delbo02, Delbo04}.
\end{enumerate}


To date the value of the thermal inertia has been derived for seven large main-belt asteroids (MBAs) \citep{1998A&A...338..340M,2004A&A...418..347M,Migo06} and six NEAs \citep{2005Icar..179...95H, 2007Icar..188..414H, Migo04, Migo07, 2004A&A...424.1075M}.  Moreover, the mean value of \ti~was estimated for the NEAs with multiwavelength thermal infrared data, the latter believed to be representative of the \TI~of NEAs with sizes between 0.8 and 3.4 km \citep{Delbo07}.

By comparing MBA and NEA \TIs, an inverse correlation between \ti~and asteroid diameter $D$  was derived \citep{Delbo07} of the form:
\begin{equation}
\Gamma = d_0 D^{-\xi},
\label{EGamma_D}
\end{equation}
{\hl where $D$ is the diameter of a sphere with a volume equivalent to that of the asteroid shape}.
Equation \ref{EGamma_D} has also important consequences for the Yarkovsky effect, implying that that the orbital semimajor axis drift rate of MBAs due to the Yarkovsky effect is proportional to $\sim$$D^{\xi-1}$ \citep{Delbo07} rather than to $D^{-1}$, the latter being the expected dependence for size independent thermal inertia. 
Given the small number of determined asteroid \TIs, \citet{Delbo07} used a unique value for $\xi$ and $d_0$ across an interval of 4 orders in magnitude in $D$. Their best-fit values are $\xi = 0.48 \pm 0.04$ and $d_0 = 300 \pm 47$, where $D$ is km and \ti~in S.I. units (\tiu).

However, there are several reasons to suspect that surface properties of large asteroids may be different from those of smaller bodies. In this case $\xi$ might acquire different values in different size ranges. For example, \citet{Bottke05} showed that asteroids with $D>$100 km and most bodies with $D>$50 km in size are likely to be primordial objects that have not suffered collisional disruption in the past 4 Gyr. These objects have resided in the asteroid belt long enough to build up a fine regolith to cause their low \ti--values regardless of their size. Moreover, the same work has shown that objects smaller than $\sim$30 km are statistically ejecta from the catastrophic collisional disruption of larger parent bodies. In the latter case, the more recent the smaller is an object. The surfaces of these asteroids might be systematically fresher with less mature and less thick regolith, implying higher--\ti~values. At the smaller end of the size distribution, an unknown role might be played by the YORP effect. By increasing the rotation rate of these bodies, regolith might have been ejected from the surfaces, leading to large \ti--values. 
Furthermore, our knowledge of \ti~for asteroids $<$20 km in size is based on NEAs only. While it is believed that NEA surfaces are representative of the small ($D<20$ km) MBA surfaces, this has still to be demonstrated. Some NEAs might have suffered planetary close approach strong enough to alter their surfaces, for instance by stripping off some of the regolith \citep[see e.g.][]{2006Icar..180..201W}. This is not the case for small MBAs. Furthermore, thermal inertia is a function of temperature \citep[\ti $\propto T^{3/2}$; see e.g.][for some discussion]{Migo07, Delbo07}. This effect may lead to \ti~offsets between cooler MBAs and hotter NEAs.

In this work we present new determination of MBA \TI~from thermophysical modeling of data obtained by the Infrared Astronomical Satellite (IRAS). We focus on MBAs with $D<100$ km in order to fill the gap of data between NEAs and the largest MBAs and improve our understating of the relation between \TI~and asteroid size.

In \S \ref{S_method} we present the method used to derive the \TI~of MBAs from IRAS data and the selection of the targets. In \S \ref{S_result} we describe the results obtained for each studied asteroid. Furthermore, in \S \ref{S_discussion}, we discuss our novel determination of asteroid \TIs~in the context of the aforementioned published results.


\section{Thermophysical modeling of IRAS data}
\label{S_method}
The Infrared Astronomical Satellites (IRAS) measured the thermal emission of more than 2200 asteroids. Asteroid thermal infrared fluxes measured by IRAS are available through the Planetary Data System on--line archives \citep{Tedesco04}. The main goal of the IRAS Minor Planet Survey \citep[IMPS;][]{Tedesco92} was the determination of asteroid sizes.
%
Due to the lack of knowledge of asteroid spin vectors and shapes, asteroid sizes of the IMPS and of its recent revision, the Supplemental IRAS Minor Planet Survey \citep[SIMPS;][]{Tedesco02}, were derived by modeling IRAS data with the ''refined'' Standard thermal model \citep[STM;][]{1986Icar...68..239L}. {\hl The STM assumes spherical, non--rotating bodies. In particular, this model assumes \ti=0, so that it can not be used to empirically constrain the thermal inertia.}
 
However, for $\sim$70 MBAs, the Asteroid Models from Lightcurve Inversion database (hereafter AMLI, a catalogue of asteroid shapes and spin vector solutions) have been made available recently\footnote{on the internet at: http://astro.troja.mff.cuni.cz/$\sim$projects/asteroids3D/web.php}. These models have been derived solving the inverse problem of determining the object's shape, its rotational state, and the scattering properties of its surface from optical lightcurves using a method developed by Mikko Kaasalainen and colleagues \citep[see][and references therein]{2002aste.conf..139K, 2001Icar..153...37K, 2001Icar..153...24K}. 

Theses asteroid shapes and spin vector solutions can be used to perform thermophysical modeling of IRAS data, thereby allowing the derivation of sizes and \TIs.

The thermal inertia of an asteroid can be derived by comparing measurements of its thermal-infrared flux to synthetic fluxes generated by means of a thermophysical model \citep[TPM;][and references therein]{Delbo04,Migo07}. A TPM uses the spin vector information to orient a mesh of planar facets (obtained from the AMLI) describing the shape of the asteroid at the time of each thermal infrared measurement. The temperature of each facet is determined by numerically solving the one-dimensional heat diffusion equation using presets \ti--values (e.g. 0, 5, 10,...,1000 \tiu). Macroscopic surface roughness is modeled by adding hemispherical craters of variable opening angle, \co, and variable surface density, \cd. Thermal conduction is also accounted for within craters. We used four preset combinations of \co~and \cd~spanning the range of possible surface roughness (see table \ref{Troughmod}).
Following the procedure of \citet{Migo07}, for each roughness model and each value of \ti, the factor $a$ that linearly scales all mesh vertices is determined by the minimization of the function {\hl
$\overline \chi^2 = 1/(N-N_f) \sum^N_{i=1} \left ( \frac{ a^2 f'_i - f_i}{\sigma_i} \right )^2$,
where $f'_i$, $f_i$, and $\sigma_i$ are the synthetic TPM generated fluxes, the IRAS thermal infrared fluxes and their quoted uncertainties, respectively; $N$ is the number of observations and $N_f$ is the number of the model parameters adjusted in the fit (degrees of freedom). In this work case $N_f$ is always equal to 2, i.e. \TI~and \De.}. 
The location of the minimum \CS~as function of \ti~gives the best--fit asteroid surface \TI~for each roughness model. The value of $a$ at \ti--minimum is used to determine the best--fit values of \De.

From the AMLI web site, we selected those MBAs with SIMPS diameters $<$100 km and at least $\sim$20 IRAS measurements. Each IRAS observation (the so--called sighting) consisted of four simultaneous measurements of the asteroid's thermal infrared flux at 12, 25, 60, and 100 $\mu m$. Our list includes (21) Lutetia, (32) Pomona, (44) Nysa, (73) Klytia, (110) Lydia, (115) Thyra, (277) Elvira, (306) Unitas, (382) Dodona, (584) Semiramis, (694) Ekard, and (720) Bohlinia. Flux values for each asteroids are reported in Table \ref{Tfluxes}. Table \ref{Tobjlist} gives basic information about the physical properties of the objects along with the number of IRAS measurements and the range of observing dates. Table \ref{Tmodels} (\SOM) report the AMLI models (donloaded in December 2007) used in this work.

\section{Results}
\label{S_result}
For each object we derived an estimate of the surface \TI~from the analysis of the plot of the \CS~as function of \ti~for different degree of surface roughness (see \S \ref{S_method})
\footnote{See the \SOM~for a detailed description of TPM results including \CS~plots for each asteroid and each spin vector solution obtained from the AMLI web site.}.
The best--fit values of \De~are given in table \ref{Tobjresults} and used in in Fig. \ref{F_tiplot} to plot \ti~vs. $D$ along with \TIs~from previous works. For those asteroids for which more than one shape and spin vector solution are available, we determined the one that gives the lowest \CS, which is the solution that we prefer.
{\hl Our results are particularly important for the asteroid (720) Bohlinia for which our preferred spin state solution was also predicted to be the corrected one just on theoretical grounds by \citet[][see also \S \ref{S_discussion}]{2003Natur.425..147V}}.

We note that, although shape uncertainties are difficult to be estimated from optical lightcurve inversion and that the constraint of convexity of the shapes from the AMLI data base plays a role in the calculation of the thermal infrared emission of these bodies, our results show that the global approximation of the shapes is in general adequate to provide a good fit of IRAS infrared measurements.
However, in the case of of (73) Klytia thermophysical modeling of IRAS data resulted in a poor fit {\hl (\CS$\sim$8; 26 data points)} independently of the spin vector solution used. We note that recent lightcurve data yield a different spin vector solutions to those reported in the AMLI (A. Carbognani, personal communication). We leave the detailed investigation of the case of (73) Klytia to a future work.

\section{Discussion}
\label{S_discussion}
This work represents the first attempt of \TI~determination of MBAs with sizes $\lesssim$100 km via thermophysical modeling of IRAS data using shapes and spin vectors derived from optical lightcurve inversion. The values derived for the \TI~are in general intermediate between those of NEAs and those obtained for the largest MBAs with sizes in the range between 200 and 1000 km.

Figure \ref{F_tiplot} shows asteroids' \TIs~derived from this work along with other values taken from the literature \citep{
Delbo07, 
2005Icar..179...95H, 2007Icar..188..414H,
Migo07, Migo06, Migo04,
1998A&A...338..340M,2004A&A...418..347M, 2004A&A...424.1075M}
plotted as function of objects' diameter. The \TI~of (54509) YORP (the leftmost data point) is a preliminary result from the study of \citet{Migo07}. 

The dashed and the dotted lines of Fig. \ref{F_tiplot} represent the fit of Eq. (\ref{EGamma_D}) to MBAs only and to NEAs only, respectively. Resulting values of $\xi$ are 1.4$\pm$0.2 for MBAs and 0.32$\pm$0.09 for NEAs. The highly different slopes derived for MBAs and NEAs indicate that a single power law gives a poor fit to the data over the $D$ range 0.1 -- 1000 km, in contrast with the results of \citet{Delbo07}, which where based on a smaller dataset of \TIvs.
%
{\hl Given the errorbars affecting asteroid \TI~determination, the \ti~vs~$D$ dependence might also be flat for $D$ in the range between 1 and 100 km and might drop for $D > 100$ km down to the low \TIv~observed for the largest bodies of the Main Belt.}
Interestingly, Fig. \ref{F_tiplot} shows that the NEA power law can reasonably fit well also MBAs with $D<$100 km (best--fit $\xi=0.21\pm0.04$ for the NEAs and the MBAs with $D<100$ km).
This might be an indication of the different regolith properties that the largest and likely primordial asteroids have in comparison to bodies with $D<$100 km, the latter probably having been catastrophically disrupted and rebuilt during the age of the solar system.

We checked the \TI~values derived by means of our method against values derived by other authors: our estimate of the \TI~of (21) Lutetia is in agreement with the \ti--values derived by \citet{Migo06}, \citet{Migo07}, and \citet{Carvano08}.  We performed thermophysical modeling of IRAS data also for some of the largest MBAs whose shape and spin vector solutions are available in the AMLI web site: for instance, we derived \ti~between 5 and 20 \tiu~for 2 Pallas. This low--\ti~value is in agreement with previous determination of the thermal inertia of this object \citep{Spencer89, 1998A&A...338..340M}.

{\hl
We note that for (720) Bohlinia the first spin solution (\lp=33.09\de, \bp=52.39\de, our preferred one) provides a better fit to IRAS data than second the spin solution (\lp=238.52\de, \bp=39.67\de). Thermophysical modeling of infrared data allowed us to discriminate between the two still possible spin state solutions obtained by the optical photometry. The first spin solution was predicted to be the correct one by \cite{2003Natur.425..147V} on theoretical grounds: they have shown that spins vectors of the four prograde-rotating
Koronis asteroids (including 720 Bohlinia) are trapped in a secular spin-orbit resonance which produces their paralelism in space. 
On the other hand, in the case of the retrograde rotator (277) Elvira, which also belongs to the Koronis family, our thermophysical analysis of IRAS data can not remove the spin solution degeneracy. However, both spin solution are theoretically possible, as the study of \cite{2003Natur.425..147V} does not put any constraint on the retrograde rotators in the Koronis family.
We note also, that due to very low inclination of the Koronis orbits any optical photometry dataset would not be able to 
distinguish between the two spin orientations, whereas the infrared data have the capability to break this degeneracy.
}

Table \ref{Tobjresults} reports the best--fit effective diameters, \De, derived by means of our TPM, for each of the studied body.
Figure \ref{F_diplot} shows the ratio between \De~and SIMPS diameters as function of \De. It can be clearly seen that \De--values tend to be larger than SIMPS diameters. 
Moreover, a correlation between the size of asteroids and the ratio between TPM and SIMPS diameters appears from Fig. \ref{F_diplot}, such that the deviation between TPM diameters and SIMPS diameters increases for smaller objects. While we caution that the data set is small, this correlation is intriguing and may be indicative of an effect due to the asteroid thermal properties: because we find that \ti~increases with decreasing asteroid size, diameters of objects derived under the STM assumption of \ti=0 are less reliable the smaller they are. It is already known that for significant \TI~the STM underestimate the real sizes. The correlation we see in Fig. \ref{F_diplot} might be due to this fact.

We leave a more detailed investigation of how SIMPS diameters compares with TPM ones and of the accuracy of the latter to a future work devoted to the derivation of sizes and thermal properties of all asteroids in the AMLI database and with IRAS data. 

We underline here the potential of our approach: it is expected that shape and spin vector solutions will be derived from optical photometry obtained for instance by the mission Gaia of the European Space Agency for more than 10,000 asteroids \citep{MignardGaiaGLO07}, or by ground based surveys such as Pan-STARRS \citep{Dur05}. Thermal infrared data will be soon available for more than 10,000 asteroids from space missions such as Spitzer, Akari, and WISE  \citep[see e.g. the work of][]{2007DPS....39.3515T}. The combination of the two data sets will enable us to use the TPMs and derive asteroid sizes and surface \TIvs down to diameters of few km in the main belt.

\section{Conclusions}
We derived the \TIvs~of 10 main belt asteroids in the size range between 30 and 100 km from thermophysical modeling of IRAS data. Our results indicate that \TI~increases with decreasing size more rapidly for main belt asteroids with diameters between 30 and 1000 km than for near-Earth asteroids smaller than 30 km. This might reflect the different regolith properties between the largest, likely primordial asteroid and the smaller ones, catastrophically disrupted and rebuilt during the age of the solar system.
We also discuss the comparison between diameters from thermophysical modeling of IRAS data and SIMPS diameters for the asteroids included in this study.
\label{S_conclusions}

\subsection*{Acknowledgments}
The work of Marco Delbo has been carried out in the framework of the European Space Agency (ESA) External Fellowship Program. 
Part of this research was also carried out while he was a Henri Poincar\'e Fellow at the Observatoire de la C\^ote d’Azur. 
The Henri Poincar\'e Fellowship is funded by the CNRS-INSU, the Conseil G\'en\'eral des Alpes-Maritimes and the Rotary International -- District 1730.\\
We thank the referees David Vokrouhlick\'y and Michael (Migo) Mueller for suggestions that led to significant improvements of the manuscript. M.D. wishes also to acknowledge fruitful discussions with Ed Tedesco. The thermophysical model was mainly run on a Opteron 8--processors computer (thot) at the \OCA~dedicated to the Gaia space mission.

{\small
\renewcommand{\baselinestretch}{0.8}

}

\newpage
\section*{Tables and Table Captions}

{
\small
\renewcommand{\baselinestretch}{0.7}

\begin{table}[h!]
\center
\begin{tabular}{|l|c|c|c|}
\hline
Model              & $\gamma_C$         & $\rho_C$ & \tb \\
\hline
  no roughness     & $0^\circ$  & 0.0 & 0$^\circ$ \\
  low roughness    & $45^\circ$ & 0.5 & 10$^\circ$ \\
  medium roughness & $68^\circ$ & 0.8 & 29$^\circ$ \\
  high roughness   & $90^\circ$ & 1.0 & 58$^\circ$\\
\hline
\end{tabular}\\
\caption{The four roughness models used in the application of the TPM to IRAS data. \tb~is the corresponding mean surface slope according to the parameterization introduced by \citet{Hapke84} \citep[see text and also][for further details]{Delbo07}}.
\label{Troughmod}
\end{table}

\begin{table}[h!]
\center

\begin{tabular}{|r|r|c|c|c|c|c|c|}
\hline
Number & Designation       & $H$    & $G$   & $p_V$  &$D$ (km)&$N_s$ &  Dates of observations  \\
\hline
   21 &            Lutetia &  7.35 & 0.11 &  0.221 & 95.760 &   20 &  1983-04-25 $\rightarrow$ 1983-05-04\\
   32 &             Pomona &  7.56 & 0.15 &  0.256 & 80.760 &   34 &  1983-07-31 $\rightarrow$ 1983-09-05\\
   44 &               Nysa &  7.03 & 0.46 &  0.546 & 70.640 &   23 &  1983-07-27 $\rightarrow$ 1983-09-01\\
   73 &             Klytia &  9.00 & 0.15 &  0.225 & 44.440 &   26 &  1983-03-10 $\rightarrow$ 1983-03-30\\
  110 &              Lydia &  7.80 & 0.20 &  0.181 & 86.090 &   20 &  1983-06-25 $\rightarrow$ 1983-07-03\\
  115 &              Thyra &  7.51 & 0.12 &  0.275 & 79.830 &   24 &  1983-04-28 $\rightarrow$ 1983-05-14\\
  277 &             Elvira &  9.84 & 0.15 &  0.277 & 27.190 &   19 &  1983-07-28 $\rightarrow$ 1983-09-01\\
  306 &             Unitas &  8.96 & 0.15 &  0.211 & 46.700 &   37 &  1983-07-31 $\rightarrow$ 1983-09-07\\
  382 &             Dodona &  8.77 & 0.15 &  0.161 & 58.370 &   21 &  1983-07-11 $\rightarrow$ 1983-08-30\\
  694 &              Ekard &  9.17 & 0.15 &  0.046 & 90.780 &   35 &  1983-06-13 $\rightarrow$ 1983-07-07\\
  720 &           Bohlinia &  9.71 & 0.15 &  0.203 & 33.730 &   18 &  1983-08-09 $\rightarrow$ 1983-09-09\\
\hline
\end{tabular}

\caption{Selected main belt asteroids with SIMPS diameters $D<$100 km, with shape and spin vector solution from lightcurve inversion and a number of IRAS sightings $N_s\geq$20. $H$ is the absolute magnitude in the $H-G$ system of \cite{Bowell89} as given in the Minor Planet Center asteroid orbits data base and $p_V$ is the SIPMS geometric visible albedo \citep{Tedesco02}. The last column reports the dates of the first and the last IRAS observations.}
\label{Tobjlist}
\end{table}


\begin{table}[h!]
\center
\begin{tabular}{|r|r|c|c|c|c|c|c|c|}
\hline
Number & Designation       & $\Gamma$&\De~(km) & $p_V$     &$D$ (km)& $p_V$      \\
       &                   & \tiu    & TPM     & TPM       &SIMPS   & SIMPS      \\
\hline                                                                             
   21 &            Lutetia &   0-100 & 107-114 & 0.16-0.18 & 96 (4) & 0.22 (0.02)\\
   32 &             Pomona &  20-120 &  84-86  & 0.22-0.24 & 81 (2) & 0.25 (0.01)\\
   44 &               Nysa &  80-160 &  80-82  & 0.40-0.42 & 71 (4) & 0.55 (0.07)\\
  110 &              Lydia &  70-200 &  90-97  & 0.14-0.16 & 86 (2) & 0.18 (0.01)\\
  115 &              Thyra &  25-100 &  90-94  & 0.20-0.22 & 80 (1) & 0.27 (0.01)\\
  277 &             Elvira & 100-400 &  36-40  & 0.13-0.16 & 27 (1) & 0.28 (0.02)\\
  306 &             Unitas & 100-260 &  55-57  & 0.14-0.15 & 47 (2) & 0.21 (0.02)\\
  382 &             Dodona &  15-150 &  74-76  & 0.095-0.10& 58 (3) & 0.16 (0.02)\\
  694 &              Ekard & 100-140 & 108-111 &0.030-0.032& 91 (4) & 0.046(0.004)\\
  720 &           Bohlinia &  70-200 &  40-42  & 0.13-0.14 & 34 (1) & 0.20 (0.02)\\
\hline
\end{tabular}

\caption{Best--fit \TI~(\ti) and effective diameters (\De) derived from TPM modeling of IRAS data. TPM $p_V$ is derived from the value of the \De~and the MPC $H$ reported in Tab.\ref{Tobjlist}. {\hl The quoted uncertainties in diameter and albedo are purely statistical.  Systematic uncertainties realted to TPM assumptions are neglected}. For comparison we list the SIMPS diameter and geometric visible albedo ($p_V$) and their quoted uncertainties within parenthesis.}
\label{Tobjresults}
\end{table}

\newpage
\section*{Figure Captions}
Figure \ref{F_tiplot}. Thermal inertia as a function of asteroid diameter. Lines with xy--errorbars represent values from the literature. $\times$ with errorbars are the results from this work (see text for details). Dotted line: fit of Eq. (\ref{EGamma_D}) to NEAs only; dashed line: fit of Eq. (\ref{EGamma_D}) to MBAs only.

Figure \ref{F_diplot}. Ratio of the diameters derived from thermophysical modeling of IRAS data of the asteroids from this work and their SIMPS diameters, plotted as function of the size of the bodies. Note the inverse correlation of the diameter ratio with size, which may be indicative of the fact the SIMPS size underestimation increases for smaller asteroids.

\newpage
\begin{figure}[h!]
\begin{center}
\includegraphics[width=10.0cm, angle=-90]{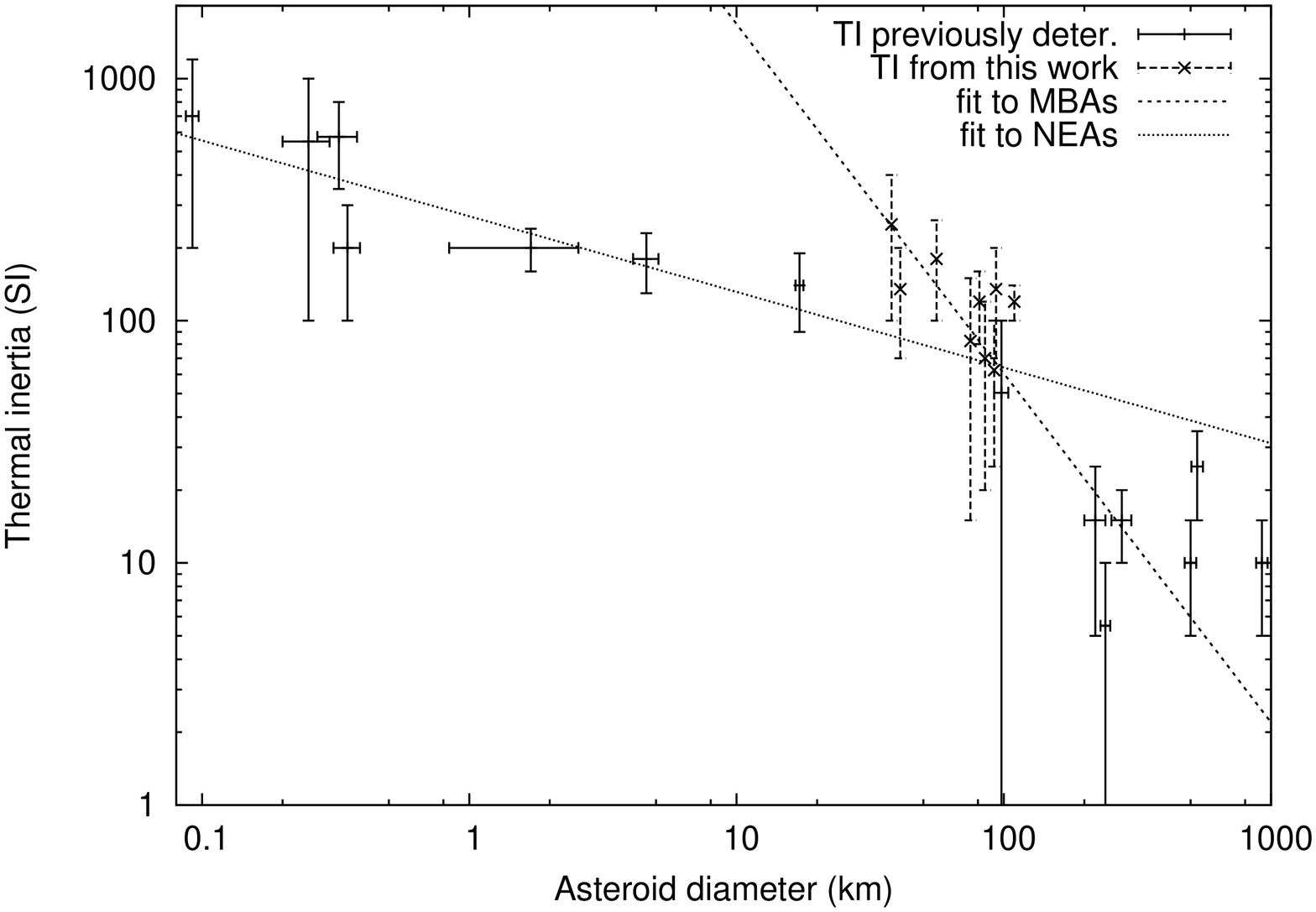}
\end{center}
\caption{~}
\label{F_tiplot}
\end{figure}

\begin{figure}[h!]
\begin{center}
\includegraphics[width=10.0cm, angle=-90]{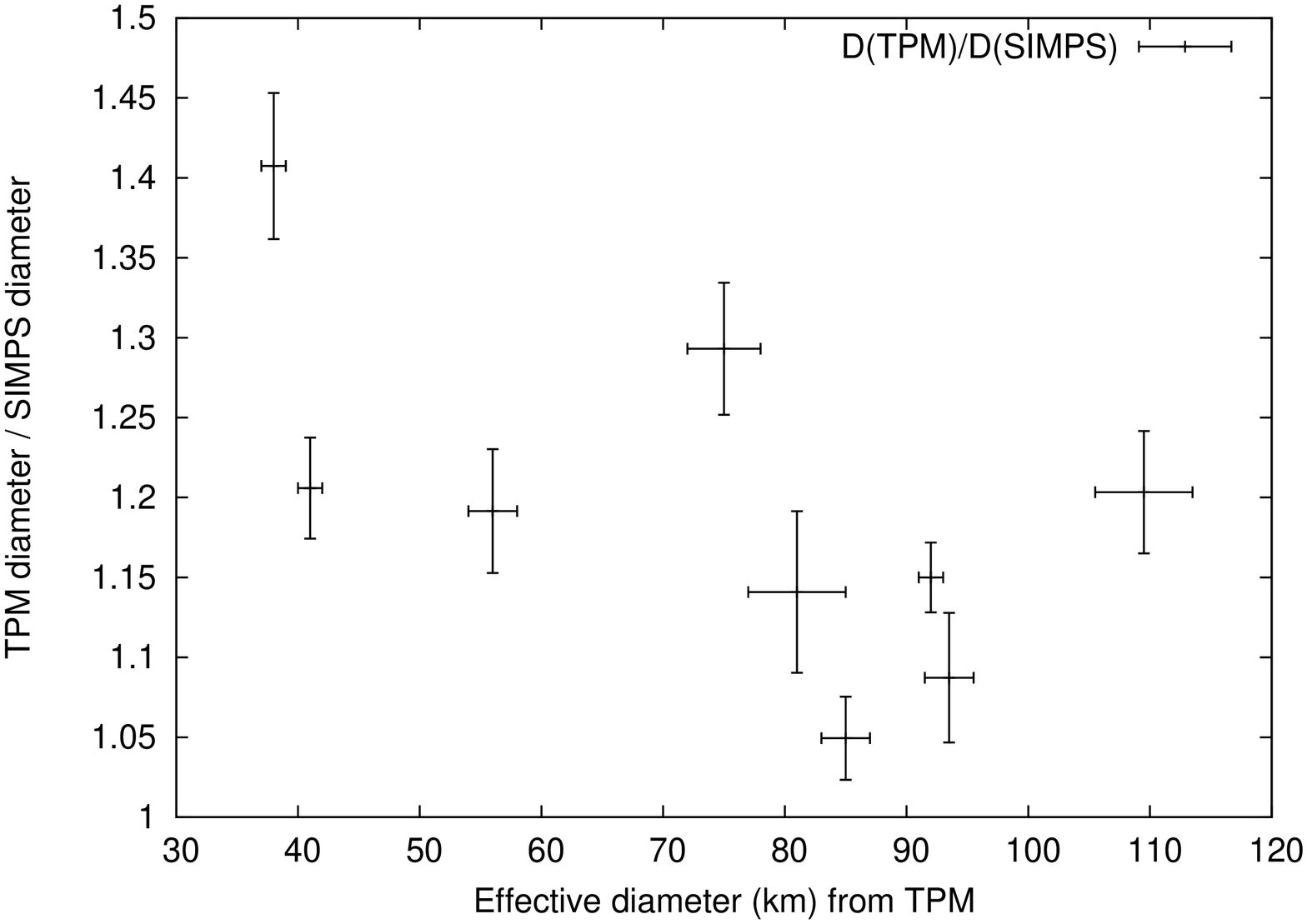}
\end{center}
\caption{~}
\label{F_diplot}
\end{figure}
}

\newpage
\section*{{\Huge \SOM}}

\section*{Description of the TPM results for each target}
\subsubsection*{(21) Lutetia}
The model \#2 of AMLI (\lp=217.77\de, \bp=12.51\de, and $P$=8.16546082 hrs) gives a lower \CS~for all roughness than the model \#1. Thermal inertia ranges from {\hl 0 to 180 \tiu. The best--fit value of \TI~is 90 \tiu.}
The corresponding best--fit \De~is between 107 and 114 km. Assuming the MPC $H$=7.35, $p_V$ is in the range between 0.156 and 0.177.\\
\subsubsection*{(32) Pomona}
A rather high degree of roughness and a {\hl \TI~between 20 and 220 \tiu~ are admissible solutions. Best fit \ti~is 112 \tiu}. The corresponding best--fit \De~is 84--86 km, that combined with the MPC $H$=7.56 yields $p_V$ of 0.22--0.24.\\
\subsubsection*{(44) Nysa}
The \TI~of this object lies in the range {\hl between 80 and 160 \tiu, with best--fit value of 115 \tiu}. The degree of surface roughness can not be constrained from the IRAS data. The corresponding best--fit \De~is between 80 and 82 km that, given then $H$ value of 7.03,  yields an albedo $p_V$ of 0.40 -- 0.42.\\
\subsubsection*{(110) Lydia}
Two spin vector and shape model solutions exist for this object. The first (\lp=149.3\de, \bp= -55.0\de, $P$=10.92580365 hrs) gives a slightly better \CS~(\CS=0.7 at \ti=95 \tiu~on the medium roughness curve) with respect to the second (\CS=0.76 at \ti=120 \tiu~on the high roughness curve). A thermal inertia between {\hl 70 and 200 \tiu~is also consistent with IRAS data}. \De~range is 90--92 km or 94--97 km depending whether the first of the second pole solution is adopted. Our choice is the first model.\\
%
%
\subsubsection*{(115) Thyra}
Roughness is not constrained for this asteroid, although a surface with a moderate to zero value of roughness is slightly preferred. Nevertheless, if the minima of all roughness model curves are included, we find that \TI~varies between {\hl 25 and 100 \tiu, with a best--fit value of 75 \tiu.} The best--fit \De~ of this asteroid is between 90 and 94 km, implying an albedo $p_V$ in the range 0.20--0.22 given the $H$ value of 7.51.\\
%
%
%
\subsubsection*{(277) Elvira}
This object has two shape and spin vector models that provide fits to the IRAS data that are almost indistinguishable. Thermal inertia ranges between {100 and 400 \tiu, with a best--fit value \ti=190 \tiu}, independent of the spin vector solution used. We note that the first model (\lp=55.99\de, \bp=-81.41\de, $P$=29.69216350 hrs) has a marginally lower \CS.
The best--fit \De~ranges from 36 to 40 km, implying a $p_V$ between 0.157 and 0.127 for $H$=9.84.\\
%
%
%
%
\subsubsection*{(306) Unitas}
Two spin vector and shape model solutions are available. The first solution provides a significantly lower \CS~than the second (the value of the \CS~drops by almost a factor of two): we adopt the first solution. The value of the best--fit \ti~ranges from 100 to about 260 \tiu, with \De~between 55 and 57 km. Assuming the MPC $H$ value of 8.98 the albedo $p_V$ of this asteroid is between 0.14 and 0.15.\\
%
%
%
\subsubsection*{(382) Dodona}
Two spin vector and shape model solutions are available. The first one gives a lower \CS~than the second. The best--fit of the second model is obtained is with no roughness: because this is unphysical, we take the first solution as our preferred one. The best--fit \TI is in the range between 15 and 150 \tiu~and the effective diameter in the range between 74 and 76 km. Assuming the MPC $H$ value of 8.77 the corresponding geometric visible albedo $p_V$ ranges between 0.095 and 0.10.\\
\subsubsection*{(694) Ekard}
Only one spin vector and shape model from the AMLI exist, and 35 IRAS sightings were acquired. Data at 12, 25, and 60 $\mu m$ have in general signal to noise ratios of 100 or more. Only the data at 100 $\mu m$ have lower signal to noise ratios, but none of these $<$10. Nevertheless, the fit of the TPM to the IRAS data is not very good, with the lowest \CS$\sim$4 on the high roughness model at \ti=140 \tiu. If the 4 data points at more than 3$\sigma$ out the TPM predictions are not included in the fit, the minimum of the \CS~drops by almost a factor of 2 (\CS~minimum $\sim$2 at \ti=140 \tiu~on the high roughness model curve). A \TI~between 100 (\CS~minimum on the medium roughness model curve) and 140 \tiu~provide the best fit to the data. A high level of surface roughness is more consistent with the IRAS data, no matter if the 4 data points at more than 3$\sigma$ from TPM predictions are included or not in the fit. The best--fit effective diameter ranges from 108 to 111 km and the corresponding value of the geometric visible albedo between 0.030 and 0.032 assuming the MPC $H$ value of 9.17, making this one of the darkest objects observed.\\
\subsubsection*{(720) Bohlinia}
This object has two shape and spin vector models. The first (\lp=33.09\de, \bp=52.39\de, $P$=8.91861864 hrs) provides a good fit if 100 $\mu m$ fluxes are removed (\CS=0.8 at \ti=100 \tiu~on the high roughness model curve and \CS=0.8 at \ti=85 \tiu~on the medium roughness model curve). For these values of \TIs~the effective diameter \De~ranges between 40 and 42 km and consequently the geometric visible albedo $p_V$ 0.13 and 0.14 assuming the MPC $H$ value of 9.71.
{\hl Note that the second spin model gives a factor 2 worse fit than the AMLI model \#1 to IRAS data, no matter whether the 100 $\mu m$ fluxes are removed or not. Note that our preferred spin model solution was predicted to be the corrected one just on theoretical
grounds \citep[][see \S \ref{S_discussion}]{2003Natur.425..147V}. \cite{2003Natur.425..147V} argued that spins of the four prograde-rotating  Koronis asteroids (including Bohlinia) is trapped in a secular spin-orbit resonance which produces their paralelism in space. Our results bring the first observational evidence that this is the case. 
Note also, that due to very low inclination of the Koronis orbits any optical photometry dataset would not be able to distinguish between the  two spin models; it is very interesting that thermophysical modeling of infrared data have the capability to break this
degeneracy.

\newpage
\section*{\SOM: Tables}
\renewcommand{\baselinestretch}{0.8}

\begin{table}[h!]
\center
\begin{tabular}{|c|c|c|c|c|c|c|}
\hline
Object  & Model & $\lambda_p$ & $\beta_p$ & $T$ (h) & $\phi_0$ & JD$_0$ \\
\hline \hline
21 Lutetia   & 1 & 52.72 & -5.54 & 8.16826946 & 0.0 & 2444822.351160\\
             & 2 & 217.77 & 12.51 & 8.16546082 & 0.0 & 2444822.351160\\ \hline
32  Pomona   & 1 & 267.07 & 57.88 & 9.44766880 & 0.0 & 2442747.264590\\ \hline
44  Nysa     & 1 & 99.22 & 57.75 & 6.42141707 & 0.0 & 2433226.633660\\ \hline
 73 Klytia   & 1 & 38.4 & +75.1 & 8.28306525 & 0.0 & 2445831.000000\\
             & 2 & 236.7 & +73.4 & 8.28306625 & 0.0 & 2445831.000000\\
             & 3 & 244.18 & +13.12 & 8.29131033 & 0.0 & 2445831.000000\\ \hline
110 Lydia    & 1 & 149.3 & -55.0 & 10.92580365 & 0.0 & 2436494.000000\\
             & 2 & 331.4 & -60.9 & 10.92580271 & 0.0 & 2436494.000000\\ \hline
115 Thyra    & 1 & 34.52 & 33.11 & 7.23996285 & 0.0 & 2443845.092510\\ \hline
277 Elvira   & 1 & 55.99 & -81.41 & 29.69216350 & 0.0 & 2445614.968300\\
             & 2 & 249.37 & -79.11 & 29.69216610 & 0.0 & 2445614.968300\\ \hline
306 Unitas   & 1 & 79.18 & -35.22 & 8.73874670 & 0.0 & 2444113.680260\\
             & 2 & 253.3 & -17.4 & 8.73874674 & 0.0 & 2444113.680260\\ \hline
382 Dodona   & 1 & 83.03 & 60.85 & 4.11322585 & 0.0 & 2445412.801460\\
             & 2 & 248.79 & 54.45 & 4.11322751 & 0.0 & 2445412.801460\\ \hline
694 Ekard    & 1 & 88.74 & -48.33 & 5.92200286 & 0.0 & 2445590.844030\\ \hline
720 Bohlinia & 1 & 33.09 & 52.39 & 8.91861864 & 0.0 & 2445467.689780\\
             & 2 & 238.52 & 39.67 & 8.91861157 & 0.0 & 2445467.689780\\
\hline
\end{tabular}

\caption{AMLI spin vector models (downloaded in December 2007) of the asteroids studied in this work. $\lambda_p$ and $\beta_p$ are the ecliptic longitude and latitude of the pole, $T$ is the rotation period of the asteroid in hours, $\phi_0$ is the absolute rotational phase of the body at the epoch JD$_0$.}
\label{Tmodels}
\end{table}

\newpage
{\tiny
\begin{center}
\begin{longtable}{|l|c|c|c|r|r|r|}
\caption[Observed IRAS fluxes and quoted uncertainties]{Observed IRAS fluxes and quoted uncertainties}\\
\hline
Object & Date & Time (UT) & JD & Wavelength & Flux (Jy) & Error (Jy) \\
       &      &           &    & ($\mu m$)  &           &            \\
\hline
\endhead
\hline
\endfoot
 (21) Lutetia    & & & & & & \\
 &  1983-04-25&14:10:23�  &  2445450.0905439816  &  12.0 	& 4.012 	& 0.398 \\
 &  1983-04-25&14:10:23�  &  2445450.0905439816  &  25.0 	& 9.989 	& 1.674 \\
 &  1983-04-25&14:10:23�  &  2445450.0905439816  &  60.0 	& 5.194 	& 1.139 \\
 &  1983-04-25&14:10:23�  &  2445450.0905439816  &  100.0 	& 2.637 	& 0.618 \\
 &  1983-04-26&00:29:13�  &  2445450.5202893517  &  12.0 	& 2.974 	& 0.393 \\
 &  1983-04-26&00:29:13�  &  2445450.5202893517  &  25.0 	& 7.808 	& 1.156 \\
 &  1983-04-26&00:29:13�  &  2445450.5202893517  &  60.0 	& 3.874 	& 0.933 \\
 &  1983-04-26&00:29:13�  &  2445450.5202893517  &  100.0 	& 2.146 	& 0.426 \\
 &  1983-04-26&02:11:59�  &  2445450.5916550928  &  12.0 	& 3.726 	& 0.430 \\
 &  1983-04-26&02:11:59�  &  2445450.5916550928  &  25.0 	& 10.100 	& 1.517 \\
 &  1983-04-26&02:11:59�  &  2445450.5916550928  &  60.0 	& 5.425 	& 1.186 \\
 &  1983-04-26&02:11:59�  &  2445450.5916550928  &  100.0 	& 1.749 	& 0.314 \\
 &  1983-05-03&14:32:31�  &  2445458.1059143520  &  12.0 	& 3.265 	& 0.327 \\
 &  1983-05-03&14:32:31�  &  2445458.1059143520  &  25.0 	& 8.813 	& 1.326 \\
 &  1983-05-03&14:32:31�  &  2445458.1059143520  &  60.0 	& 4.158 	& 0.902 \\
 &  1983-05-03&14:32:31�  &  2445458.1059143520  &  100.0 	& 2.119 	& 0.506 \\
 &  1983-05-04&00:50:44�  &  2445458.5352314813  &  12.0 	& 3.159 	& 0.377 \\
 &  1983-05-04&00:50:44�  &  2445458.5352314813  &  25.0 	& 9.399 	& 1.574 \\
 &  1983-05-04&00:50:44�  &  2445458.5352314813  &  60.0 	& 3.219 	& 0.783 \\
 &  1983-05-04&00:50:44�  &  2445458.5352314813  &  100.0 	& 1.568 	& 0.338 \\
 (32) Pomona     & & & & & & \\
 &  1983-07-31&01:09:28�  &  2445546.5482407408  &  12.0 	& 2.178 	& 0.268 \\
 &  1983-07-31&01:09:28�  &  2445546.5482407408  &  25.0 	& 6.243 	& 0.897 \\
 &  1983-07-31&01:09:28�  &  2445546.5482407408  &  60.0 	& 3.561 	& 0.842 \\
 &  1983-07-31&02:53:41�  &  2445546.6206134260  &  12.0 	& 1.974 	& 0.260 \\
 &  1983-07-31&02:53:41�  &  2445546.6206134260  &  25.0 	& 4.857 	& 0.778 \\
 &  1983-07-31&02:53:41�  &  2445546.6206134260  &  60.0 	& 2.659 	& 0.572 \\
 &  1983-07-31&13:16:15�  &  2445547.0529513890  &  12.0 	& 2.386 	& 0.287 \\
 &  1983-07-31&13:16:15�  &  2445547.0529513890  &  25.0 	& 6.022 	& 0.927 \\
 &  1983-07-31&13:16:15�  &  2445547.0529513890  &  60.0 	& 3.354 	& 0.798 \\
 &  1983-07-31&13:16:15�  &  2445547.0529513890  &  100.0 	& 1.399 	& 0.253 \\
 &  1983-08-03&21:58:56�  &  2445550.4159259261  &  12.0 	& 2.458 	& 0.271 \\
 &  1983-08-03&21:58:56�  &  2445550.4159259261  &  25.0 	& 7.156 	& 1.208 \\
 &  1983-08-03&21:58:56�  &  2445550.4159259261  &  60.0 	& 2.407 	& 0.551 \\
 &  1983-08-03&21:58:56�  &  2445550.4159259261  &  100.0 	& 1.482 	& 0.249 \\
 &  1983-08-03&23:41:13�  &  2445550.4869560185  &  12.0 	& 2.423 	& 0.286 \\
 &  1983-08-03&23:41:13�  &  2445550.4869560185  &  25.0 	& 6.723 	& 1.050 \\
 &  1983-08-03&23:41:13�  &  2445550.4869560185  &  60.0 	& 3.575 	& 0.848 \\
 &  1983-08-03&23:41:13�  &  2445550.4869560185  &  100.0 	& 1.085 	& 0.233 \\
 &  1983-08-11&20:36:42�  &  2445558.3588194447  &  12.0 	& 2.507 	& 0.286 \\
 &  1983-08-11&20:36:42�  &  2445558.3588194447  &  25.0 	& 6.752 	& 1.104 \\
 &  1983-08-11&20:36:42�  &  2445558.3588194447  &  60.0 	& 3.274 	& 0.769 \\
 &  1983-08-11&20:36:42�  &  2445558.3588194447  &  100.0 	& 1.258 	& 0.252 \\
 &  1983-08-19&15:47:32�  &  2445566.1580092590  &  12.0 	& 3.004 	& 0.322 \\
 &  1983-08-19&15:47:32�  &  2445566.1580092590  &  25.0 	& 7.544 	& 1.132 \\
 &  1983-08-19&15:47:32�  &  2445566.1580092590  &  60.0 	& 4.746 	& 1.130 \\
 &  1983-08-19&15:47:32�  &  2445566.1580092590  &  100.0 	& 1.361 	& 0.266 \\
 &  1983-09-05&20:06:39�  &  2445583.3379513887  &  12.0 	& 3.951 	& 0.394 \\
 &  1983-09-05&20:06:39�  &  2445583.3379513887  &  25.0 	& 10.239 	& 1.521 \\
 &  1983-09-05&20:06:39�  &  2445583.3379513887  &  60.0 	& 4.776 	& 1.038 \\
 &  1983-09-05&20:06:39�  &  2445583.3379513887  &  100.0 	& 1.771 	& 0.273 \\
 &  1983-09-05&18:24:29�  &  2445583.2670023148  &  12.0 	& 3.394 	& 0.353 \\
 &  1983-09-05&18:24:29�  &  2445583.2670023148  &  25.0 	& 8.766 	& 1.331 \\
 &  1983-09-05&18:24:29�  &  2445583.2670023148  &  60.0 	& 5.049 	& 1.101 \\
 &  1983-09-05&18:24:29�  &  2445583.2670023148  &  100.0 	& 1.834 	& 0.314 \\
 (44) Nysa       & & & & & & \\
 &  1983-07-27&19:54:32�  &  2445543.3295370368  &  12.0 	& 2.199 	& 0.329 \\
 &  1983-07-27&19:54:32�  &  2445543.3295370368  &  25.0 	& 7.315 	& 1.168 \\
 &  1983-07-27&19:54:32�  &  2445543.3295370368  &  60.0 	& 2.706 	& 0.636 \\
 &  1983-07-27&21:37:44�  &  2445543.4012037036  &  12.0 	& 2.820 	& 0.364 \\
 &  1983-07-27&21:37:44�  &  2445543.4012037036  &  25.0 	& 8.113 	& 1.151 \\
 &  1983-07-27&21:37:44�  &  2445543.4012037036  &  60.0 	& 3.782 	& 0.818 \\
 &  1983-07-27&21:37:44�  &  2445543.4012037036  &  100.0 	& 1.782 	& 0.319 \\
 &  1983-08-08&18:41:22�  &  2445555.2787268520  &  12.0 	& 3.415 	& 0.340 \\
 &  1983-08-08&18:41:22�  &  2445555.2787268520  &  25.0 	& 9.318 	& 1.571 \\
 &  1983-08-08&18:41:22�  &  2445555.2787268520  &  60.0 	& 4.324 	& 1.029 \\
 &  1983-08-08&18:41:22�  &  2445555.2787268520  &  100.0 	& 2.041 	& 0.381 \\
 &  1983-08-08&20:24:25�  &  2445555.3502893518  &  12.0 	& 2.175 	& 0.289 \\
 &  1983-08-08&20:24:25�  &  2445555.3502893518  &  25.0 	& 6.083 	& 1.073 \\
 &  1983-08-08&20:24:25�  &  2445555.3502893518  &  60.0 	& 3.075 	& 0.664 \\
 &  1983-08-08&20:24:25�  &  2445555.3502893518  &  100.0 	& 1.164 	& 0.250 \\
 &  1983-09-01&14:48:14�  &  2445579.1168287038  &  12.0 	& 4.854 	& 0.540 \\
 &  1983-09-01&14:48:14�  &  2445579.1168287038  &  25.0 	& 13.025 	& 1.798 \\
 &  1983-09-01&14:48:14�  &  2445579.1168287038  &  60.0 	& 6.260 	& 1.376 \\
 &  1983-09-01&14:48:14�  &  2445579.1168287038  &  100.0 	& 2.280 	& 0.395 \\
 &  1983-09-01&16:27:21�  &  2445579.1856597224  &  12.0 	& 4.001 	& 0.466 \\
 &  1983-09-01&16:27:21�  &  2445579.1856597224  &  25.0 	& 11.252 	& 1.658 \\
 &  1983-09-01&16:27:21�  &  2445579.1856597224  &  60.0 	& 5.458 	& 1.316 \\
 &  1983-09-01&16:27:21�  &  2445579.1856597224  &  100.0 	& 2.147 	& 0.371 \\
(110) Lydia      & & & & & & \\
 &  1983-06-25&13:58:32�  &  2445511.0823148149  &  12.0 	& 2.607 	& 0.373 \\
 &  1983-06-25&13:58:32�  &  2445511.0823148149  &  25.0 	& 7.050 	& 1.053 \\
 &  1983-06-25&13:58:32�  &  2445511.0823148149  &  60.0 	& 3.989 	& 0.941 \\
 &  1983-06-25&13:58:32�  &  2445511.0823148149  &  100.0 	& 0.920 	& 0.188 \\
 &  1983-06-25&10:32:48�  &  2445510.9394444446  &  12.0 	& 2.503 	& 0.302 \\
 &  1983-06-25&10:32:48�  &  2445510.9394444446  &  25.0 	& 7.412 	& 1.107 \\
 &  1983-06-25&10:32:48�  &  2445510.9394444446  &  60.0 	& 4.152 	& 0.996 \\
 &  1983-06-25&10:32:48�  &  2445510.9394444446  &  100.0 	& 1.683 	& 0.311 \\
 &  1983-06-25&12:15:38�  &  2445511.0108564813  &  12.0 	& 2.325 	& 0.292 \\
 &  1983-06-25&12:15:38�  &  2445511.0108564813  &  25.0 	& 6.511 	& 1.083 \\
 &  1983-06-25&12:15:38�  &  2445511.0108564813  &  60.0 	& 3.002 	& 0.644 \\
 &  1983-06-25&12:15:38�  &  2445511.0108564813  &  100.0 	& 1.576 	& 0.279 \\
 &  1983-07-03&10:54:18�  &  2445518.9543750002  &  12.0 	& 2.545 	& 0.438 \\
 &  1983-07-03&10:54:18�  &  2445518.9543750002  &  25.0 	& 6.946 	& 1.002 \\
 &  1983-07-03&10:54:18�  &  2445518.9543750002  &  60.0 	& 3.604 	& 0.860 \\
 &  1983-07-03&10:54:18�  &  2445518.9543750002  &  100.0 	& 1.287 	& 0.258 \\
 &  1983-07-03&12:37:23�  &  2445519.0259606480  &  12.0 	& 2.674 	& 0.331 \\
 &  1983-07-03&12:37:23�  &  2445519.0259606480  &  25.0 	& 5.515 	& 0.878 \\
 &  1983-07-03&12:37:23�  &  2445519.0259606480  &  60.0 	& 3.285 	& 0.703 \\
 &  1983-07-03&12:37:23�  &  2445519.0259606480  &  100.0 	& 1.366 	& 0.270 \\
(115) Thyra      & & & & & & \\
 &  1983-04-28&02:17:23�  &  2445452.5954050925  &  12.0 	& 4.881 	& 0.542 \\
 &  1983-04-28&02:17:23�  &  2445452.5954050925  &  25.0 	& 9.941 	& 1.676 \\
 &  1983-04-28&02:17:23�  &  2445452.5954050925  &  60.0 	& 4.753 	& 1.034 \\
 &  1983-04-28&02:17:23�  &  2445452.5954050925  &  100.0 	& 1.793 	& 0.369 \\
 &  1983-04-28&04:00:52�  &  2445452.6672685184  &  12.0 	& 4.926 	& 0.566 \\
 &  1983-04-28&04:00:52�  &  2445452.6672685184  &  25.0 	& 11.379 	& 1.686 \\
 &  1983-04-28&04:00:52�  &  2445452.6672685184  &  60.0 	& 5.302 	& 1.163 \\
 &  1983-04-28&04:00:52�  &  2445452.6672685184  &  100.0 	& 2.662 	& 0.591 \\
 &  1983-05-06&07:54:25�  &  2445460.8294560187  &  12.0 	& 4.572 	& 0.455 \\
 &  1983-05-06&07:54:25�  &  2445460.8294560187  &  25.0 	& 9.601 	& 1.431 \\
 &  1983-05-06&07:54:25�  &  2445460.8294560187  &  60.0 	& 4.690 	& 1.021 \\
 &  1983-05-06&07:54:25�  &  2445460.8294560187  &  100.0 	& 1.328 	& 0.244 \\
 &  1983-05-06&09:37:40�  &  2445460.9011574076  &  12.0 	& 3.922 	& 0.393 \\
 &  1983-05-06&09:37:40�  &  2445460.9011574076  &  25.0 	& 9.040 	& 1.507 \\
 &  1983-05-06&09:37:40�  &  2445460.9011574076  &  60.0 	& 4.785 	& 1.045 \\
 &  1983-05-06&09:37:40�  &  2445460.9011574076  &  100.0 	& 1.433 	& 0.297 \\
 &  1983-05-14&11:43:54�  &  2445468.9888194446  &  12.0 	& 3.949 	& 0.421 \\
 &  1983-05-14&11:43:54�  &  2445468.9888194446  &  25.0 	& 8.612 	& 1.412 \\
 &  1983-05-14&11:43:54�  &  2445468.9888194446  &  60.0 	& 3.801 	& 0.827 \\
 &  1983-05-14&11:43:54�  &  2445468.9888194446  &  100.0 	& 1.769 	& 0.368 \\
 &  1983-05-14&13:27:14�  &  2445469.0605787039  &  12.0 	& 3.506 	& 0.352 \\
 &  1983-05-14&13:27:14�  &  2445469.0605787039  &  25.0 	& 7.789 	& 1.317 \\
 &  1983-05-14&13:27:14�  &  2445469.0605787039  &  60.0 	& 3.926 	& 0.926 \\
 &  1983-05-14&13:27:14�  &  2445469.0605787039  &  100.0 	& 1.911 	& 0.407 \\
(277) Elvira     & & & & & & \\
 &  1983-07-28&13:13:19�  &  2445544.0509143518  &  12.0 	& 0.436 	& 0.082 \\
 &  1983-07-28&13:13:19�  &  2445544.0509143518  &  25.0 	& 1.144 	& 0.231 \\
 &  1983-07-28&13:13:19�  &  2445544.0509143518  &  60.0 	& 0.669 	& 0.145 \\
 &  1983-07-28&13:13:19�  &  2445544.0509143518  &  100.0 	& 1.446 	& 0.316 \\
 &  1983-07-28&14:49:54�  &  2445544.1179861110  &  12.0 	& 0.507 	& 0.093 \\
 &  1983-07-28&14:49:54�  &  2445544.1179861110  &  25.0 	& 0.939 	& 0.215 \\
 &  1983-07-28&14:49:54�  &  2445544.1179861110  &  60.0 	& 0.593 	& 0.112 \\
 &  1983-08-08&23:51:54�  &  2445555.4943749998  &  12.0 	& 0.447 	& 0.082 \\
 &  1983-08-08&23:51:54�  &  2445555.4943749998  &  25.0 	& 0.948 	& 0.196 \\
 &  1983-08-08&23:51:54�  &  2445555.4943749998  &  60.0 	& 0.561 	& 0.106 \\
 &  1983-08-09&01:34:53�  &  2445555.5658912039  &  12.0 	& 0.352 	& 0.068 \\
 &  1983-08-09&01:34:53�  &  2445555.5658912039  &  25.0 	& 0.849 	& 0.175 \\
 &  1983-08-09&01:34:53�  &  2445555.5658912039  &  60.0 	& 0.637 	& 0.136 \\
 &  1983-09-01&14:49:41�  &  2445579.1178356484  &  12.0 	& 0.541 	& 0.090 \\
 &  1983-09-01&14:49:41�  &  2445579.1178356484  &  25.0 	& 1.543 	& 0.282 \\
 &  1983-09-01&14:49:41�  &  2445579.1178356484  &  60.0 	& 0.819 	& 0.157 \\
 &  1983-09-01&16:28:48�  &  2445579.1866666665  &  12.0 	& 0.615 	& 0.098 \\
 &  1983-09-01&16:28:48�  &  2445579.1866666665  &  25.0 	& 1.702 	& 0.304 \\
 &  1983-09-01&16:28:48�  &  2445579.1866666665  &  60.0 	& 0.753 	& 0.149 \\
(306) Unitas     & & & & & & \\
 &  1983-07-31&02:51:32�  &  2445546.6191203706  &  12.0 	& 3.490 	& 0.441 \\
 &  1983-07-31&02:51:32�  &  2445546.6191203706  &  25.0 	& 5.988 	& 0.883 \\
 &  1983-07-31&02:51:32�  &  2445546.6191203706  &  60.0 	& 2.619 	& 0.619 \\
 &  1983-07-31&01:07:19�  &  2445546.5467476854  &  12.0 	& 2.521 	& 0.321 \\
 &  1983-07-31&01:07:19�  &  2445546.5467476854  &  25.0 	& 4.683 	& 0.747 \\
 &  1983-07-31&01:07:19�  &  2445546.5467476854  &  60.0 	& 2.331 	& 0.552 \\
 &  1983-07-31&13:14:06�  &  2445547.0514583332  &  12.0 	& 2.328 	& 0.327 \\
 &  1983-07-31&13:14:06�  &  2445547.0514583332  &  25.0 	& 4.753 	& 0.673 \\
 &  1983-07-31&13:14:06�  &  2445547.0514583332  &  60.0 	& 2.004 	& 0.473 \\
 &  1983-08-04&01:21:10�  &  2445550.5563657410  &  12.0 	& 2.214 	& 0.259 \\
 &  1983-08-04&01:21:10�  &  2445550.5563657410  &  25.0 	& 4.760 	& 0.666 \\
 &  1983-08-04&01:21:10�  &  2445550.5563657410  &  60.0 	& 2.378 	& 0.562 \\
 &  1983-08-03&23:39:01�  &  2445550.4854282406  &  12.0 	& 2.998 	& 0.310 \\
 &  1983-08-03&23:39:01�  &  2445550.4854282406  &  25.0 	& 5.569 	& 0.787 \\
 &  1983-08-03&23:39:01�  &  2445550.4854282406  &  60.0 	& 2.657 	& 0.628 \\
 &  1983-08-12&15:28:44�  &  2445559.1449537035  &  12.0 	& 3.057 	& 0.305 \\
 &  1983-08-12&15:28:44�  &  2445559.1449537035  &  25.0 	& 5.737 	& 0.930 \\
 &  1983-08-12&15:28:44�  &  2445559.1449537035  &  60.0 	& 2.640 	& 0.566 \\
 &  1983-08-12&13:47:29�  &  2445559.0746412035  &  12.0 	& 2.761 	& 0.312 \\
 &  1983-08-12&13:47:29�  &  2445559.0746412035  &  25.0 	& 5.133 	& 0.827 \\
 &  1983-08-12&13:47:29�  &  2445559.0746412035  &  60.0 	& 2.574 	& 0.548 \\
 &  1983-08-20&15:48:60�  &  2445567.1590277776  &  12.0 	& 3.495 	& 0.348 \\
 &  1983-08-20&15:48:60�  &  2445567.1590277776  &  25.0 	& 7.034 	& 1.050 \\
 &  1983-08-20&15:48:60�  &  2445567.1590277776  &  60.0 	& 3.350 	& 0.788 \\
 &  1983-08-20&15:48:60�  &  2445567.1590277776  &  100.0 	& 1.882 	& 0.318 \\
 &  1983-08-20&17:32:12�  &  2445567.2306944444  &  12.0 	& 3.166 	& 0.348 \\
 &  1983-08-20&17:32:12�  &  2445567.2306944444  &  25.0 	& 6.159 	& 1.008 \\
 &  1983-08-20&17:32:12�  &  2445567.2306944444  &  60.0 	& 2.718 	& 0.581 \\
 &  1983-08-20&17:32:12�  &  2445567.2306944444  &  100.0 	& 2.199 	& 0.430 \\
 &  1983-09-07&21:53:22�  &  2445585.4120601853  &  12.0 	& 4.294 	& 0.466 \\
 &  1983-09-07&21:53:22�  &  2445585.4120601853  &  25.0 	& 9.710 	& 1.596 \\
 &  1983-09-07&21:53:22�  &  2445585.4120601853  &  60.0 	& 2.987 	& 0.707 \\
 &  1983-09-07&21:53:22�  &  2445585.4120601853  &  100.0 	& 2.218 	& 0.417 \\
 &  1983-09-07&23:36:27�  &  2445585.4836458336  &  12.0 	& 3.929 	& 0.391 \\
 &  1983-09-07&23:36:27�  &  2445585.4836458336  &  25.0 	& 8.472 	& 1.247 \\
 &  1983-09-07&23:36:27�  &  2445585.4836458336  &  60.0 	& 3.968 	& 0.942 \\
 &  1983-09-07&23:36:27�  &  2445585.4836458336  &  100.0 	& 2.588 	& 0.511 \\
(382) Dodona     & & & & & & \\
 &  1983-07-11&11:21:47�  &  2445526.9734606482  &  12.0 	& 1.554 	& 0.208 \\
 &  1983-07-11&11:21:47�  &  2445526.9734606482  &  25.0 	& 4.083 	& 0.570 \\
 &  1983-07-11&11:21:47�  &  2445526.9734606482  &  60.0 	& 2.515 	& 0.594 \\
 &  1983-07-11&11:21:47�  &  2445526.9734606482  &  100.0 	& 1.002 	& 0.204 \\
 &  1983-07-11&13:04:52�  &  2445527.0450462964  &  12.0 	& 2.173 	& 0.261 \\
 &  1983-07-11&13:04:52�  &  2445527.0450462964  &  25.0 	& 5.039 	& 0.706 \\
 &  1983-07-11&13:04:52�  &  2445527.0450462964  &  60.0 	& 2.316 	& 0.495 \\
 &  1983-07-11&13:04:52�  &  2445527.0450462964  &  100.0 	& 1.481 	& 0.333 \\
 &  1983-07-23&08:32:23�  &  2445538.8558217594  &  12.0 	& 2.092 	& 0.319 \\
 &  1983-07-23&08:32:23�  &  2445538.8558217594  &  25.0 	& 5.110 	& 0.798 \\
 &  1983-07-23&08:32:23�  &  2445538.8558217594  &  60.0 	& 2.815 	& 0.669 \\
 &  1983-07-23&10:15:39�  &  2445538.9275347223  &  12.0 	& 2.051 	& 0.290 \\
 &  1983-07-23&10:15:39�  &  2445538.9275347223  &  25.0 	& 5.334 	& 0.854 \\
 &  1983-07-23&10:15:39�  &  2445538.9275347223  &  60.0 	& 1.927 	& 0.453 \\
 &  1983-07-23&10:15:39�  &  2445538.9275347223  &  100.0 	& 1.059 	& 0.193 \\
 &  1983-08-30&06:59:28�  &  2445576.7912962963  &  12.0 	& 1.222 	& 0.179 \\
 &  1983-08-30&06:59:28�  &  2445576.7912962963  &  25.0 	& 3.274 	& 0.543 \\
 &  1983-08-30&06:59:28�  &  2445576.7912962963  &  60.0 	& 1.506 	& 0.297 \\
 &  1983-08-30&08:42:35�  &  2445576.8629050925  &  12.0 	& 1.276 	& 0.201 \\
 &  1983-08-30&08:42:35�  &  2445576.8629050925  &  25.0 	& 3.165 	& 0.553 \\
 &  1983-08-30&08:42:35�  &  2445576.8629050925  &  60.0 	& 1.650 	& 0.341 \\
(694) Ekard      & & & & & & \\
 &  1983-06-13&22:52:16�  &  2445499.4529629629  &  12.0 	& 24.364 	& 2.842 \\
 &  1983-06-13&22:52:16�  &  2445499.4529629629  &  25.0 	& 32.217 	& 3.317 \\
 &  1983-06-13&22:52:16�  &  2445499.4529629629  &  60.0 	& 13.672 	& 2.824 \\
 &  1983-06-13&22:52:16�  &  2445499.4529629629  &  100.0 	& 3.947 	& 0.858 \\
 &  1983-06-14&00:34:56�  &  2445499.5242592595  &  12.0 	& 26.906 	& 3.291 \\
 &  1983-06-14&00:34:56�  &  2445499.5242592595  &  25.0 	& 43.510 	& 4.489 \\
 &  1983-06-14&00:34:56�  &  2445499.5242592595  &  60.0 	& 15.685 	& 3.743 \\
 &  1983-06-14&00:34:56�  &  2445499.5242592595  &  100.0 	& 6.844 	& 1.497 \\
 &  1983-06-13&21:09:33�  &  2445499.3816319443  &  12.0 	& 25.625 	& 3.058 \\
 &  1983-06-13&21:09:33�  &  2445499.3816319443  &  25.0 	& 38.516 	& 3.970 \\
 &  1983-06-13&21:09:33�  &  2445499.3816319443  &  60.0 	& 18.138 	& 4.659 \\
 &  1983-06-13&21:09:33�  &  2445499.3816319443  &  100.0 	& 4.553 	& 0.914 \\
 &  1983-06-23&14:48:24�  &  2445509.1169444444  &  12.0 	& 23.688 	& 2.914 \\
 &  1983-06-23&14:48:24�  &  2445509.1169444444  &  25.0 	& 35.971 	& 3.991 \\
 &  1983-06-23&14:48:24�  &  2445509.1169444444  &  60.0 	& 15.918 	& 3.319 \\
 &  1983-06-23&14:48:24�  &  2445509.1169444444  &  100.0 	& 4.604 	& 1.127 \\
 &  1983-06-23&16:31:40�  &  2445509.1886574072  &  12.0 	& 27.770 	& 3.704 \\
 &  1983-06-23&16:31:40�  &  2445509.1886574072  &  25.0 	& 38.734 	& 3.992 \\
 &  1983-06-23&16:31:40�  &  2445509.1886574072  &  60.0 	& 18.129 	& 4.385 \\
 &  1983-06-23&16:31:40�  &  2445509.1886574072  &  100.0 	& 5.715 	& 1.265 \\
 &  1983-06-23&18:14:53�  &  2445509.2603356480  &  12.0 	& 30.152 	& 3.589 \\
 &  1983-06-23&18:14:53�  &  2445509.2603356480  &  25.0 	& 47.105 	& 4.862 \\
 &  1983-06-23&18:14:53�  &  2445509.2603356480  &  60.0 	& 17.334 	& 3.990 \\
 &  1983-06-23&18:14:53�  &  2445509.2603356480  &  100.0 	& 6.954 	& 1.698 \\
 &  1983-07-07&15:29:12�  &  2445523.1452777777  &  12.0 	& 31.737 	& 3.140 \\
 &  1983-07-07&15:29:12�  &  2445523.1452777777  &  25.0 	& 48.717 	& 5.029 \\
 &  1983-07-07&15:29:12�  &  2445523.1452777777  &  60.0 	& 17.682 	& 4.059 \\
 &  1983-07-07&15:29:12�  &  2445523.1452777777  &  100.0 	& 7.714 	& 1.889 \\
 &  1983-07-07&05:17:52�  &  2445522.7207407407  &  12.0 	& 17.712 	& 1.904 \\
 &  1983-07-07&05:17:52�  &  2445522.7207407407  &  25.0 	& 55.794 	& 5.766 \\
 &  1983-07-07&05:17:52�  &  2445522.7207407407  &  60.0 	& 23.059 	& 6.503 \\
 &  1983-07-07&05:17:52�  &  2445522.7207407407  &  100.0 	& 9.138 	& 2.025 \\
 &  1983-07-07&13:45:56�  &  2445523.0735648149  &  12.0 	& 33.854 	& 4.502 \\
 &  1983-07-07&13:45:56�  &  2445523.0735648149  &  25.0 	& 56.635 	& 5.853 \\
 &  1983-07-07&13:45:56�  &  2445523.0735648149  &  100.0 	& 7.475 	& 1.607 \\
(720) Bohlinia   & & & & & & \\
 &  1983-08-09&07:41:37�  &  2445555.8205671296  &  12.0 	& 0.562 	& 0.111 \\
 &  1983-08-09&07:41:37�  &  2445555.8205671296  &  25.0 	& 1.520 	& 0.329 \\
 &  1983-08-09&07:41:37�  &  2445555.8205671296  &  60.0 	& 0.445 	& 0.089 \\
 &  1983-08-09&05:58:41�  &  2445555.7490856480  &  12.0 	& 0.349 	& 0.058 \\
 &  1983-08-09&05:58:41�  &  2445555.7490856480  &  25.0 	& 1.249 	& 0.254 \\
 &  1983-08-09&05:58:41�  &  2445555.7490856480  &  60.0 	& 0.672 	& 0.124 \\
 &  1983-08-09&05:58:41�  &  2445555.7490856480  &  100.0 	& 1.280 	& 0.223 \\
 &  1983-08-21&03:03:08�  &  2445567.6271759258  &  12.0 	& 0.429 	& 0.072 \\
 &  1983-08-21&03:03:08�  &  2445567.6271759258  &  25.0 	& 1.268 	& 0.292 \\
 &  1983-08-21&03:03:08�  &  2445567.6271759258  &  60.0 	& 0.547 	& 0.114 \\
 &  1983-08-21&06:29:16�  &  2445567.7703240742  &  25.0 	& 1.147 	& 0.230 \\
 &  1983-08-21&06:29:16�  &  2445567.7703240742  &  60.0 	& 0.581 	& 0.108 \\
 &  1983-08-21&04:46:11�  &  2445567.6987384260  &  12.0 	& 0.380 	& 0.062 \\
 &  1983-08-21&04:46:11�  &  2445567.6987384260  &  25.0 	& 1.056 	& 0.234 \\
 &  1983-08-21&04:46:11�  &  2445567.6987384260  &  60.0 	& 0.572 	& 0.104 \\
 &  1983-09-09&10:51:47�  &  2445586.9526273147  &  12.0 	& 0.309 	& 0.053 \\
 &  1983-09-09&10:51:47�  &  2445586.9526273147  &  25.0 	& 1.227 	& 0.281 \\
 &  1983-09-09&10:51:47�  &  2445586.9526273147  &  60.0 	& 0.473 	& 0.088 \\
\hline
\label{Tfluxes}
\end{longtable}
\end{center}

}

\newpage
{
\small
\renewcommand{\baselinestretch}{0.7}

\newpage
\section*{\SOM: Figures and Figure Captions}

\begin{figure}[h!]
\begin{center}
\includegraphics[width=8.0cm, angle=-90]{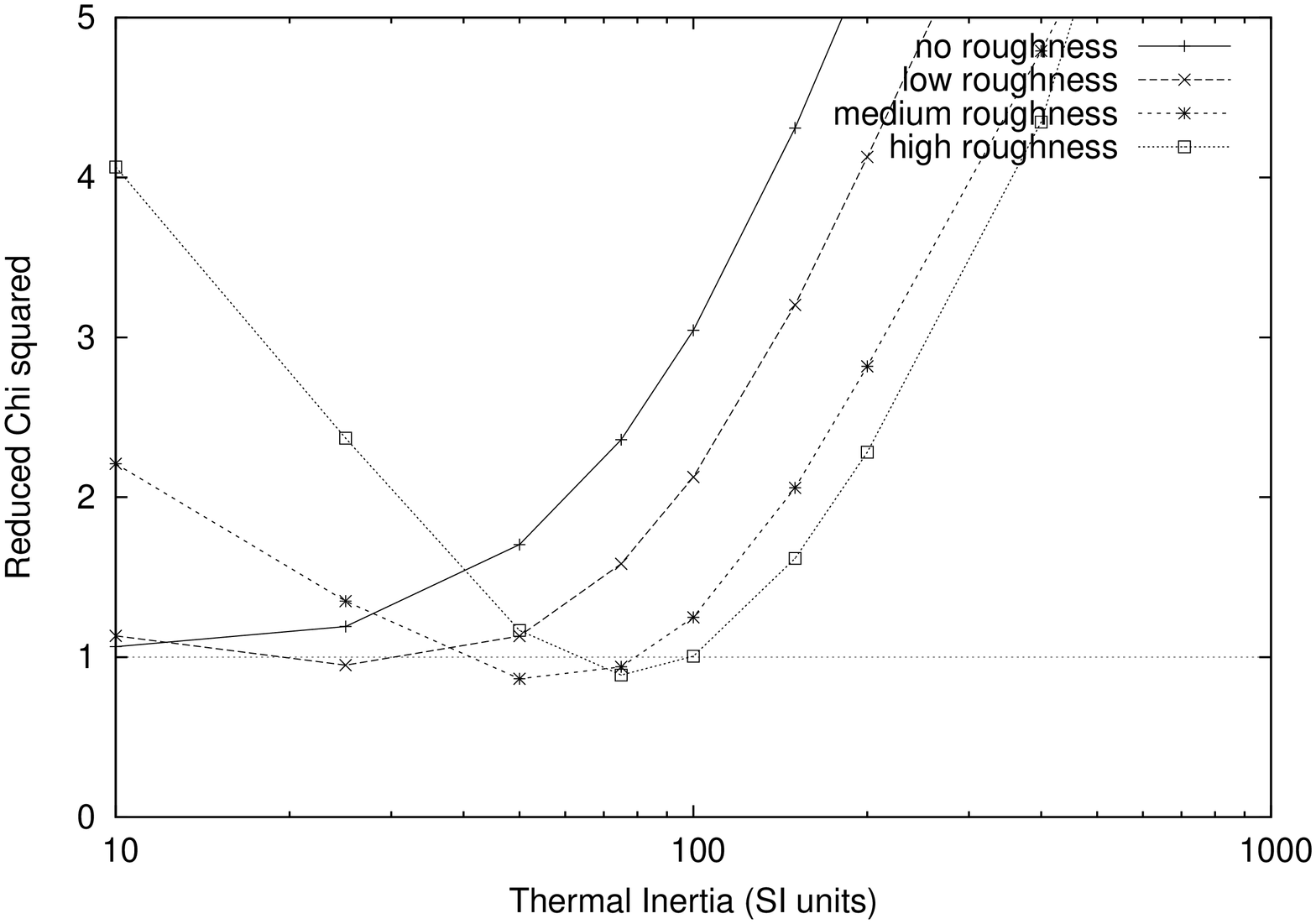}
\includegraphics[width=8.0cm, angle=-90]{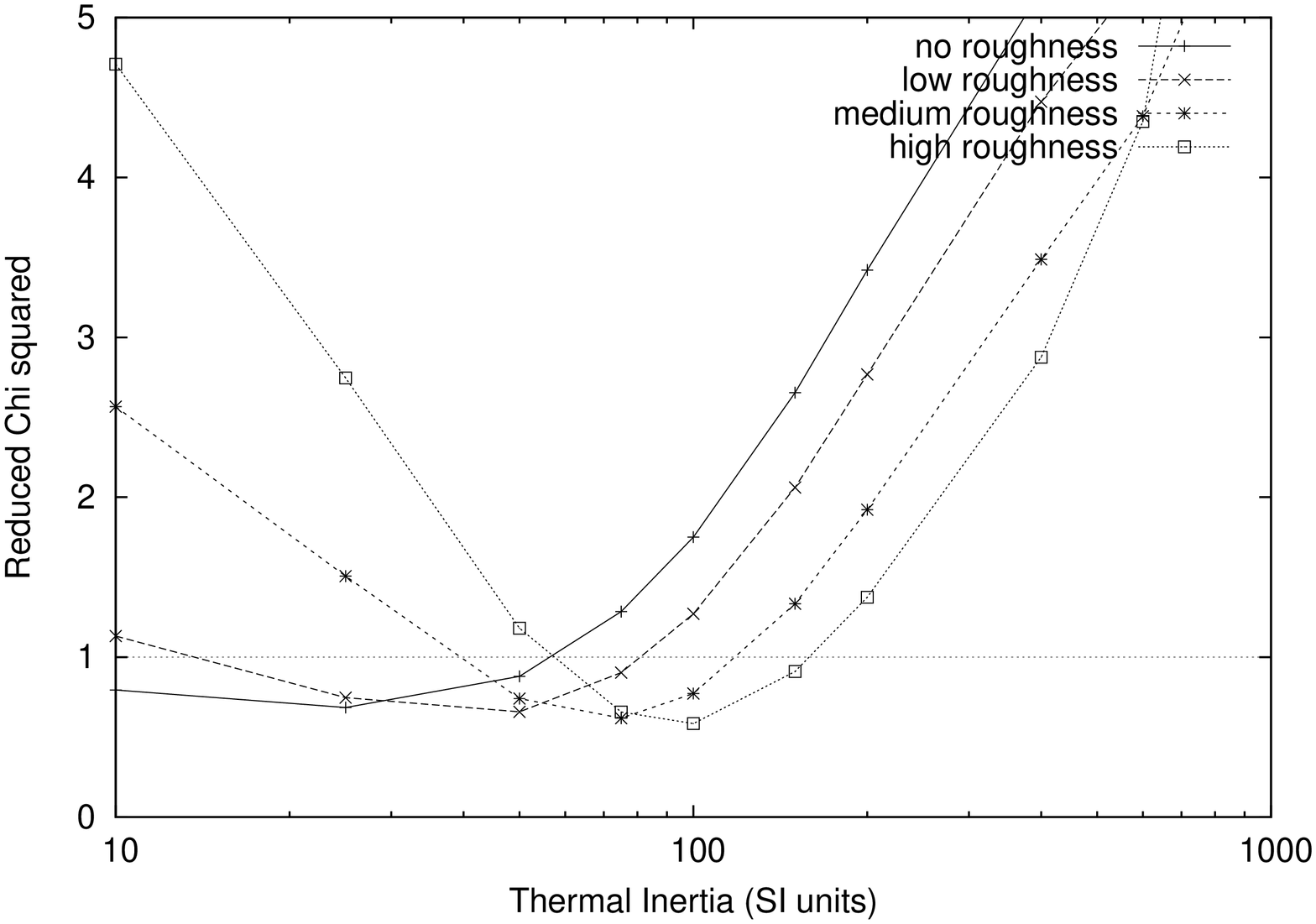}
\end{center}
\caption{{\hl Reduced \CS~of the TPM fit to IRAS infrared data as function of the \TI~for the asteroid (21) Lutetia. See \S \ref{S_method} for the definition of the \CS~adopted in this work}. Each curve corresponds to a different roughness model (see the legend on the top right of the plot and Table \ref{Troughmod}). The best--fit \TI~is the abscissa of the minimum \CS. {\hl An horizontal line is drawn at \CS=1. Admissible solution values for thermal inertia and surface roughness are defined by that portion of the curves with \CS $\leq$ 1}. \MMP}
\label{F21}
\end{figure}

\begin{figure}[h!]
\begin{center}
\includegraphics[width=8.0 cm, angle=-90]{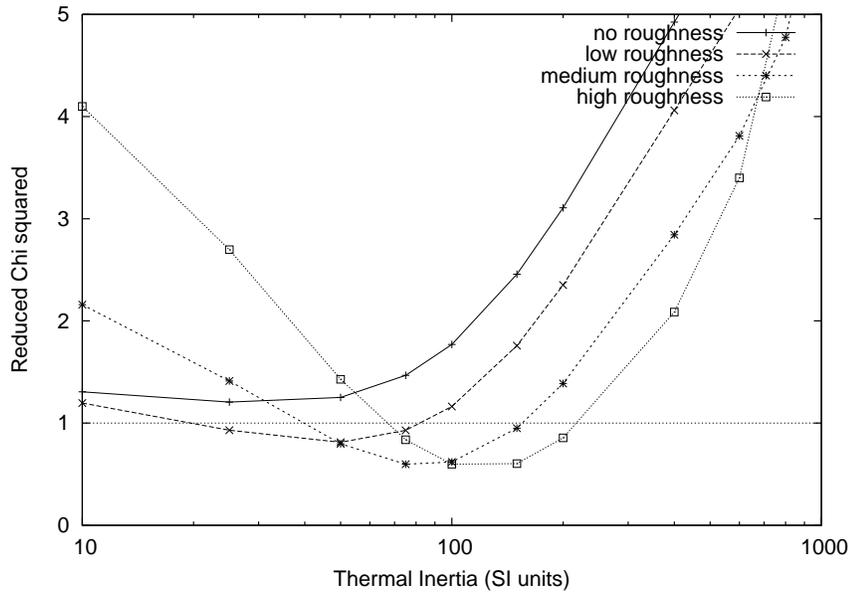}
\end{center}
\caption{As of Fig. \ref{F21} but for the asteroid (32) Pomona.}
\label{F32}
\end{figure}

\begin{figure}[h!]
\begin{center}
\includegraphics[width=8.0cm, angle=-90]{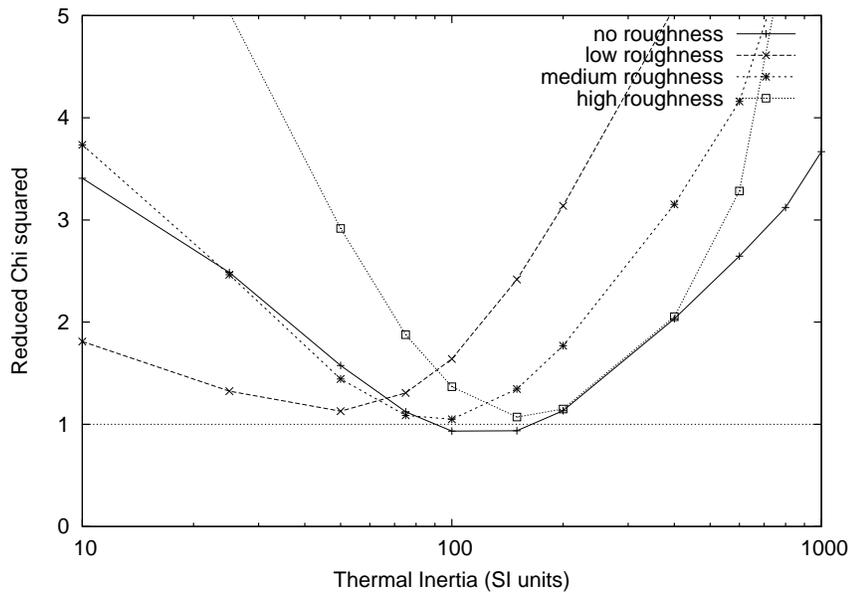}
\end{center}
\caption{As of Fig. \ref{F21} but for the asteroid (44) Nysa.}
\label{F44}
\end{figure}

\begin{figure}[h!]
\begin{center}
\includegraphics[width=8.0cm, angle=-90]{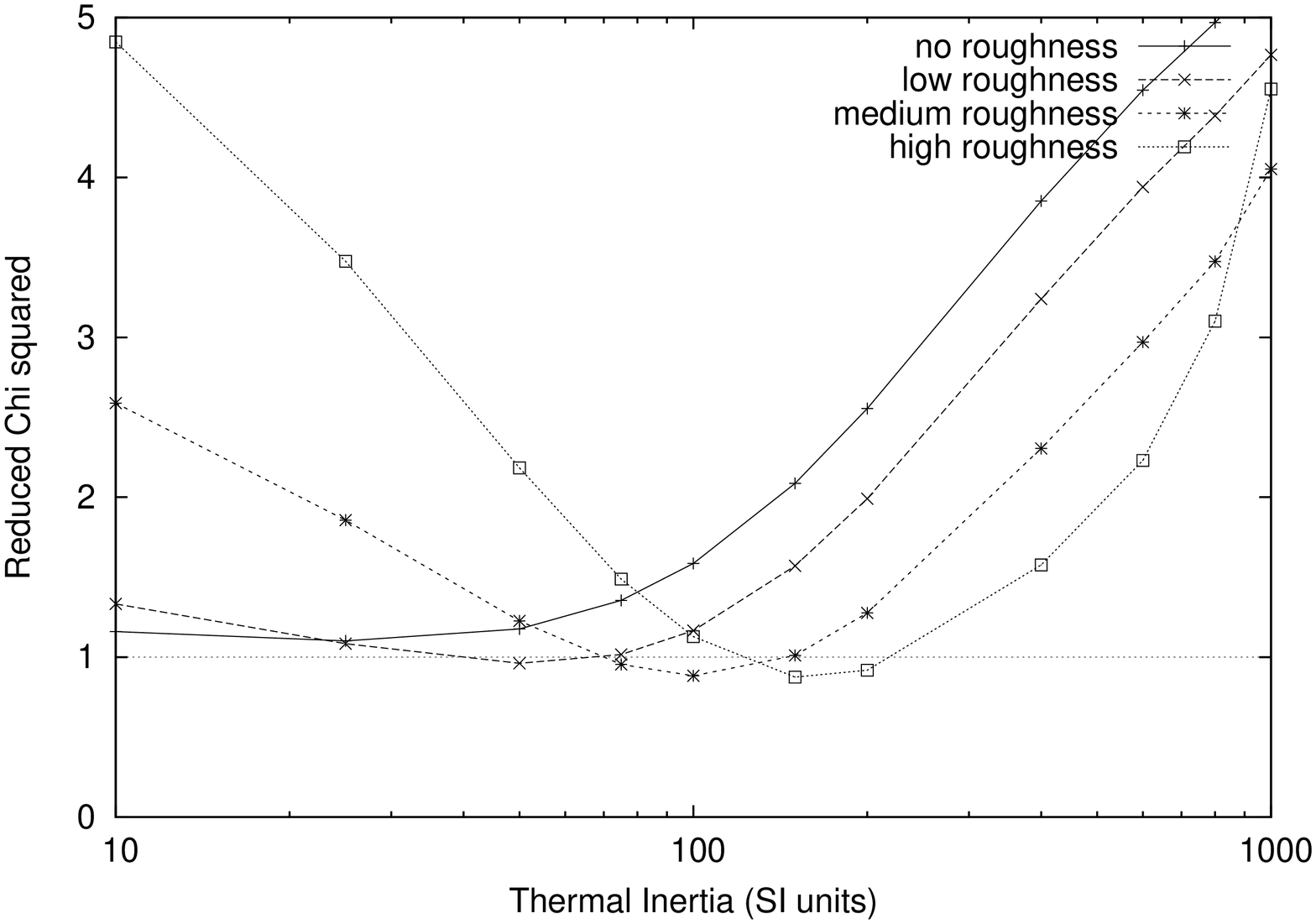}
\includegraphics[width=8.0cm, angle=-90]{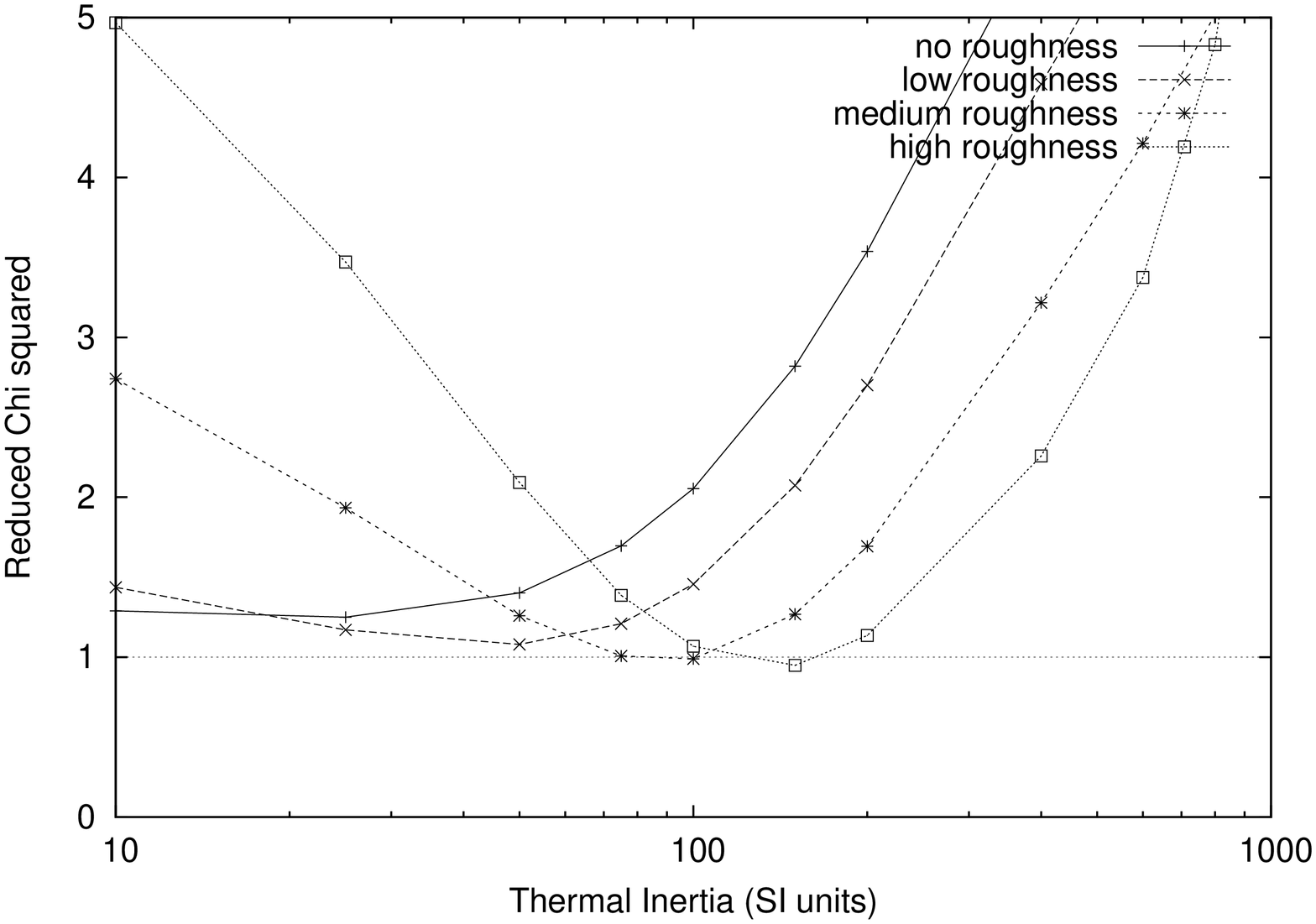}
\end{center}
\caption{As of Fig. \ref{F21} but for the asteroid (110) Lydia. \MMP}
\label{F110}
\end{figure}

\begin{figure}[h!]
\begin{center}
\includegraphics[width=8.0cm, angle=-90]{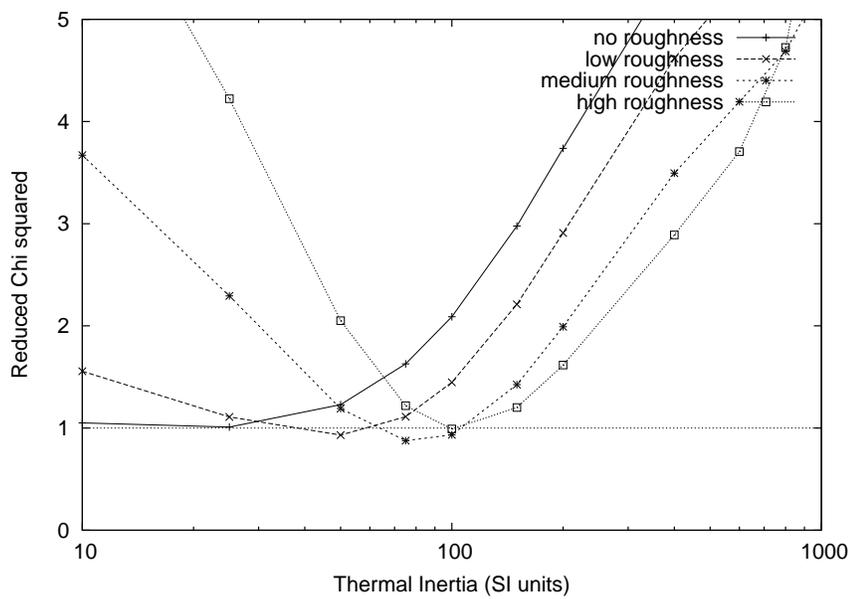}
\end{center}
\caption{As of Fig. \ref{F21} but for the asteroid (115) Thyra.}
\label{F115}
\end{figure}

\begin{figure}[h!]
\begin{center}
\includegraphics[width=8.0cm, angle=-90]{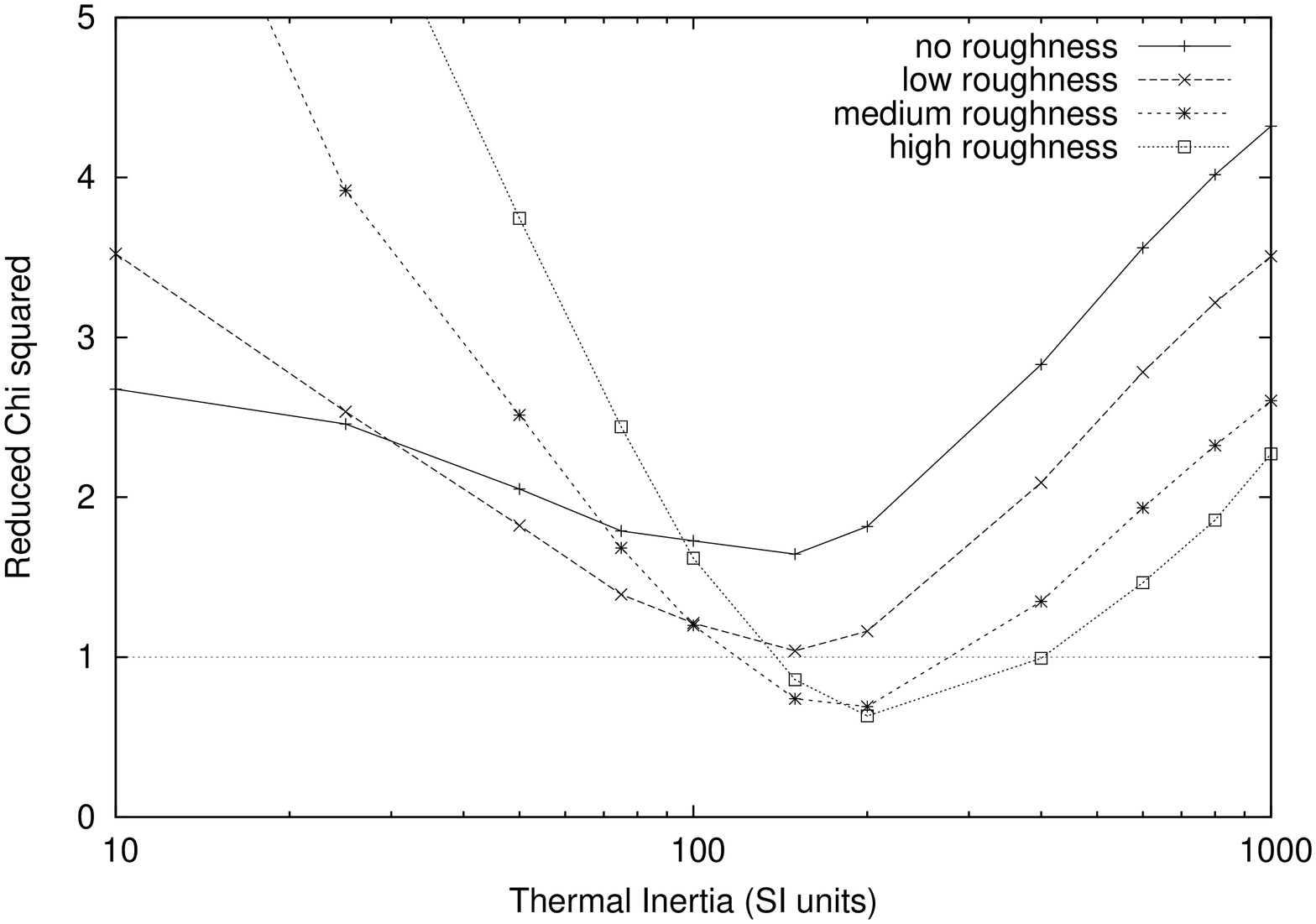}
\includegraphics[width=8.0cm, angle=-90]{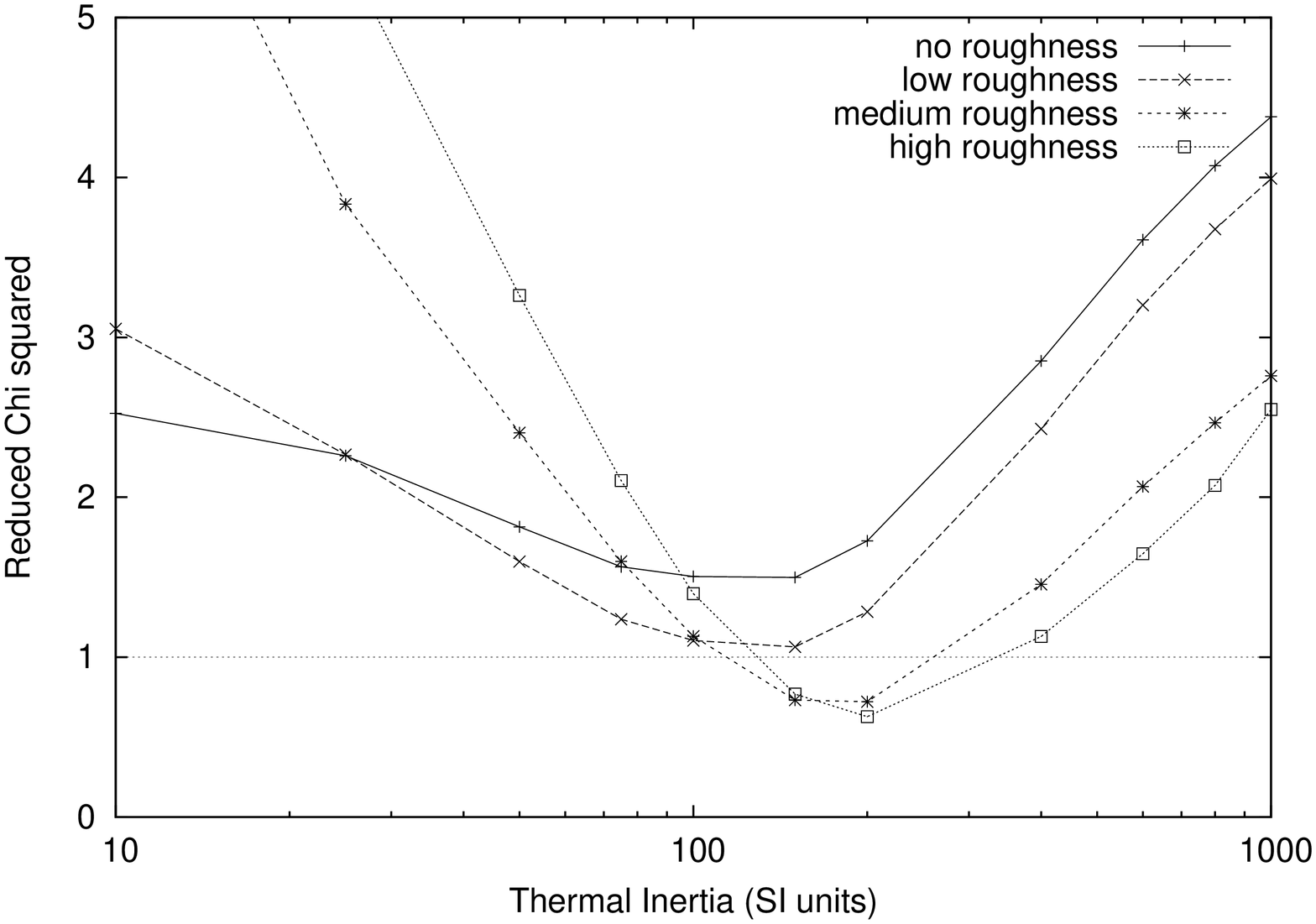}
\end{center}
\caption{As of Fig. \ref{F21} but for the asteroid (277) Elvira. \MMP.}
\label{F277}
\end{figure}

\begin{figure}[h!]
\begin{center}
\includegraphics[width=8.0cm, angle=-90]{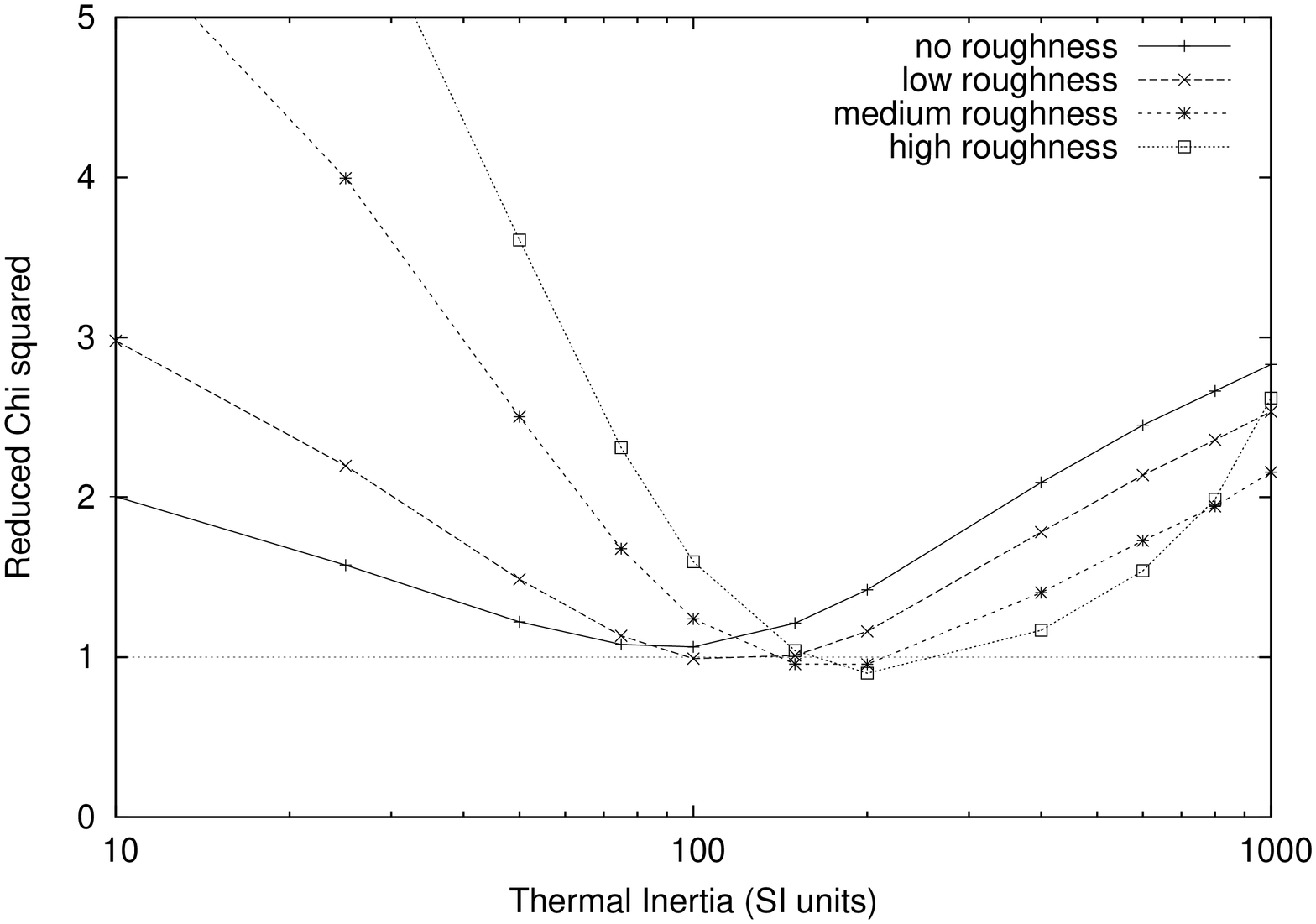}
\includegraphics[width=8.0cm, angle=-90]{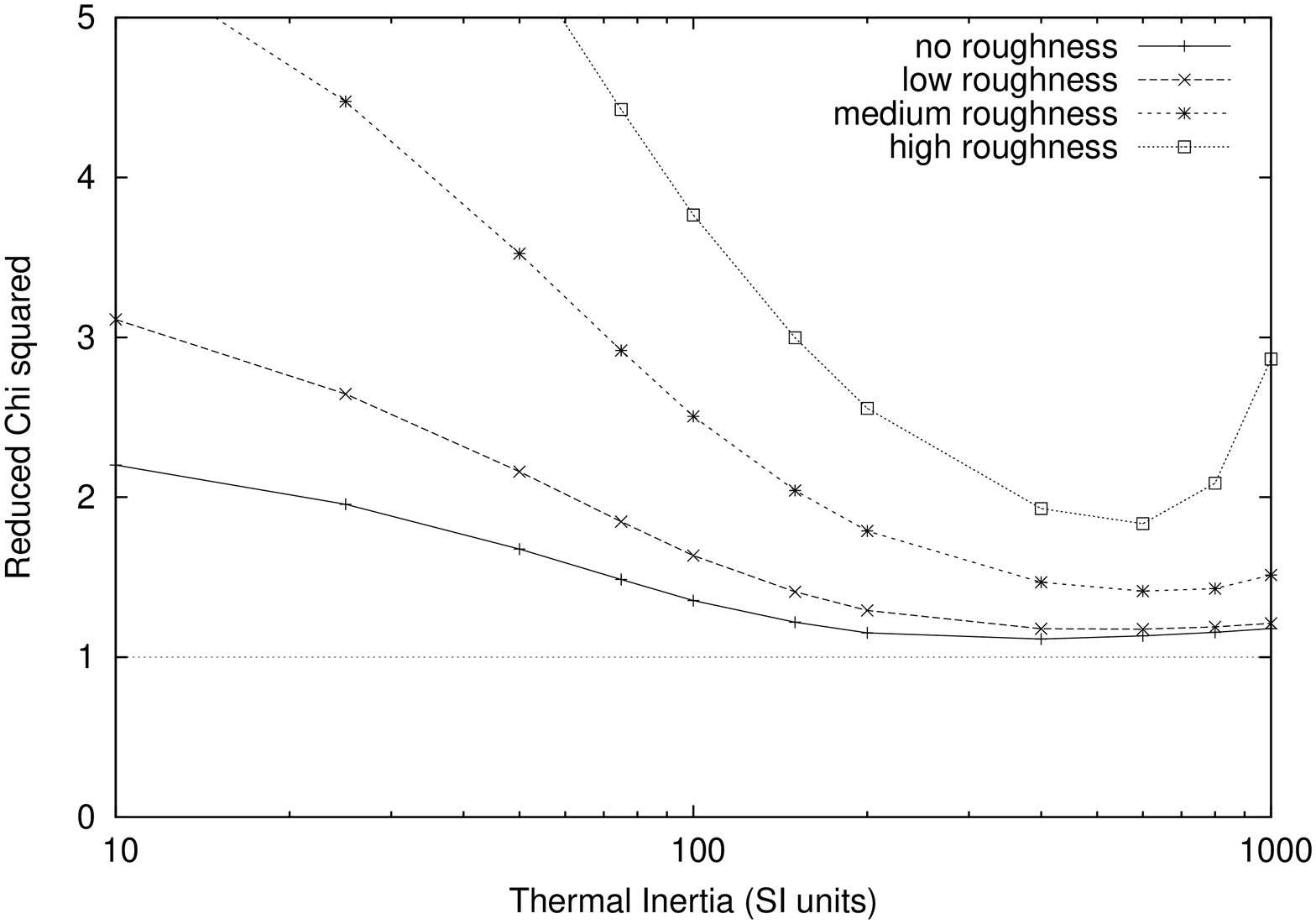}
\end{center}
\caption{As of Fig. \ref{F21} but for the asteroid (306) Unitas. \MMP.}
\label{F306}
\end{figure}

\begin{figure}[h!]
\begin{center}
\includegraphics[width=8.0cm, angle=-90]{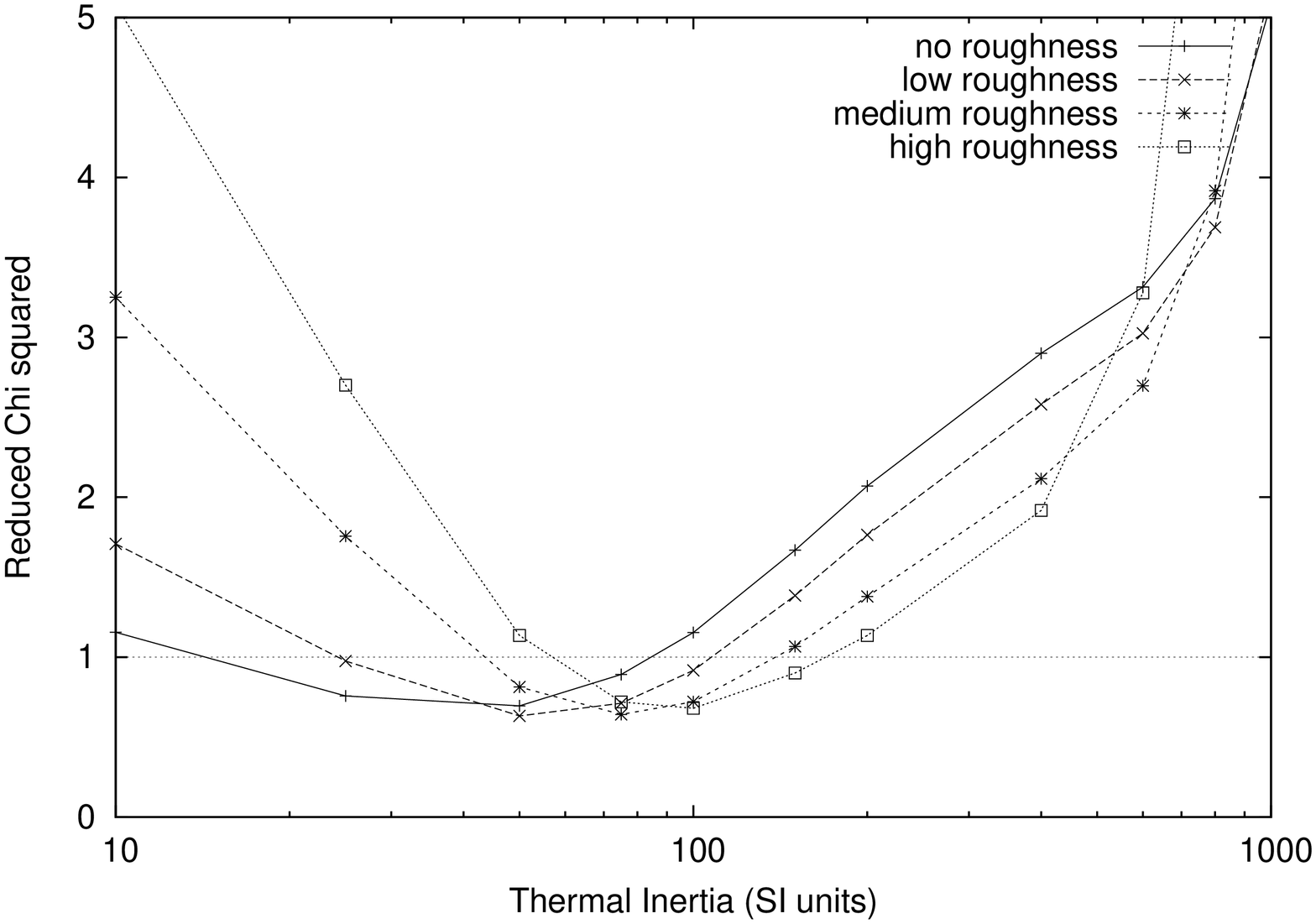}
\includegraphics[width=8.0cm, angle=-90]{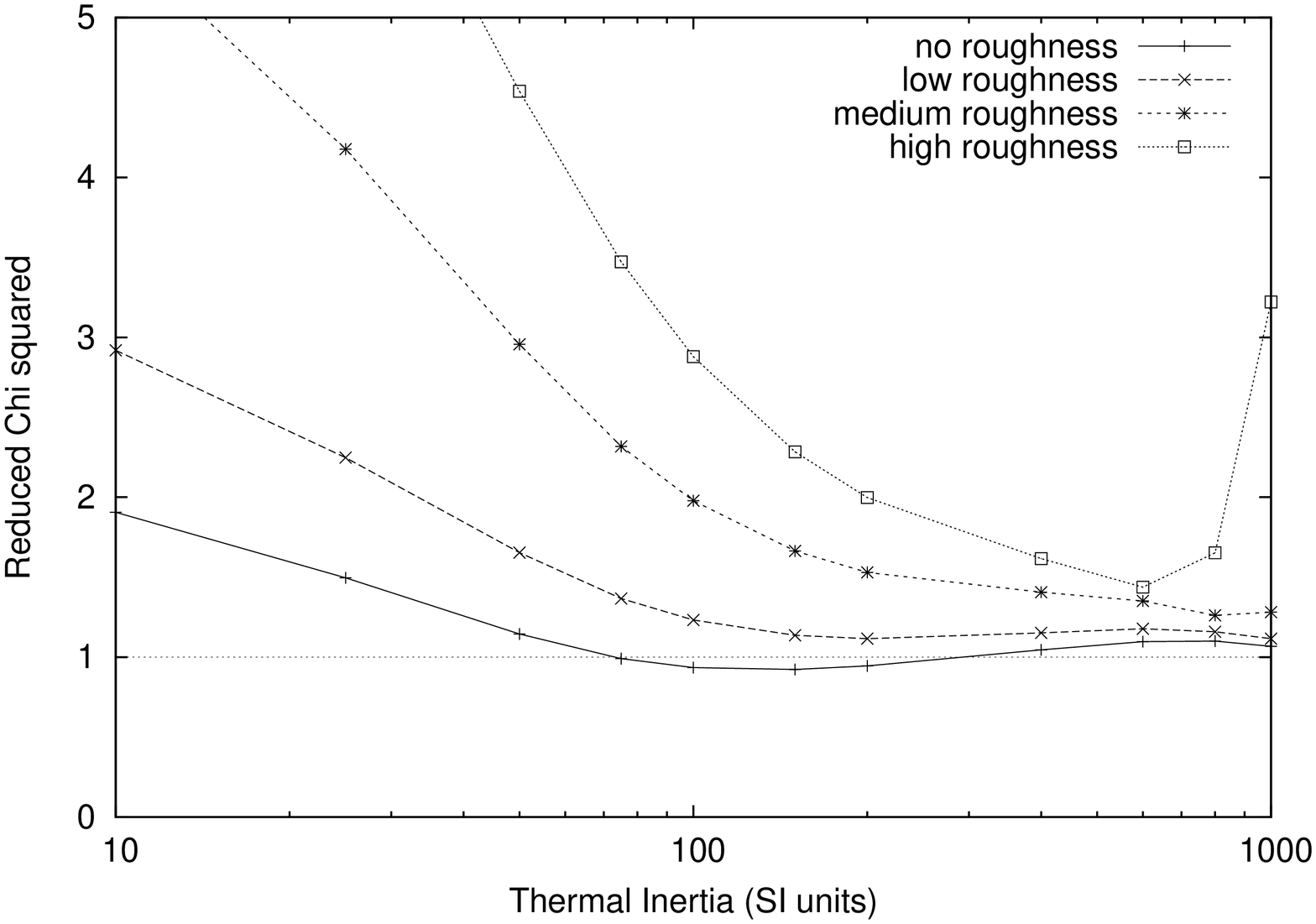}
\end{center}
\caption{As of Fig. \ref{F21} but for the asteroid (382) Dodona. \MMP.}
\end{figure}

\begin{figure}[h!]
\begin{center}
\includegraphics[width=8.0cm, angle=-90]{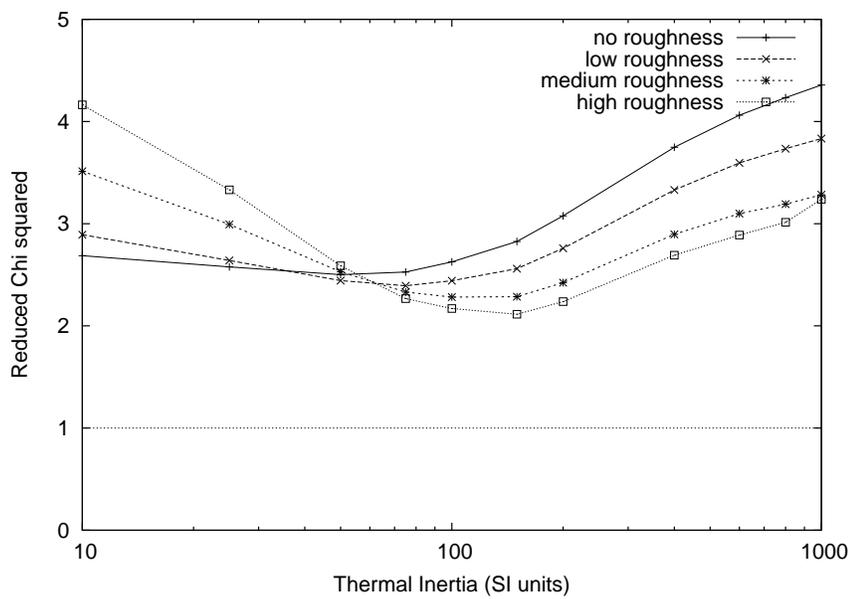}
\end{center}
\caption{As of Fig. \ref{F21} but for the asteroid (694) Ekard.}
\label{F694}
\end{figure}

\begin{figure}[h!]
\begin{center}
\includegraphics[width=8.0cm, angle=-90]{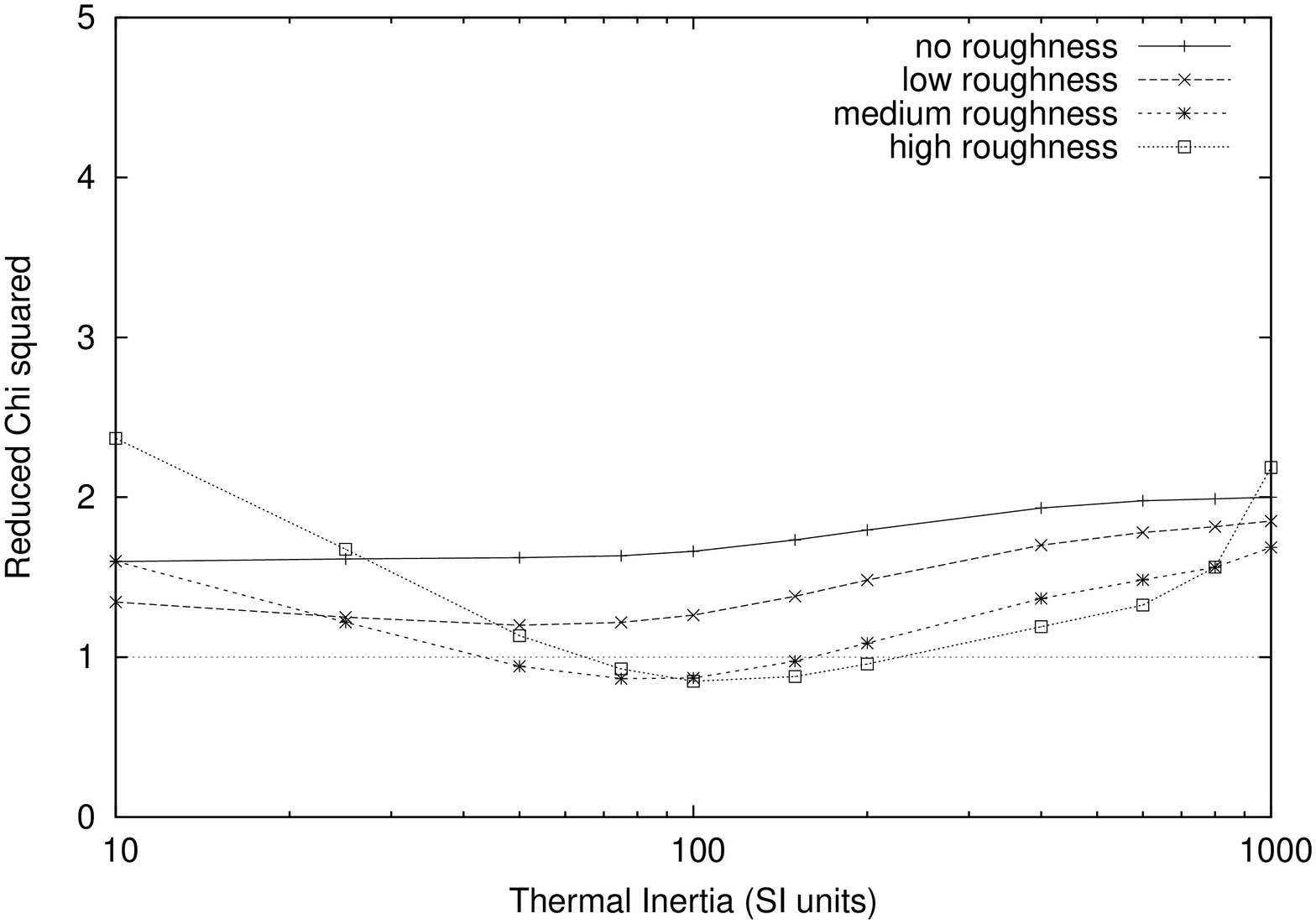}
\includegraphics[width=8.0cm, angle=-90]{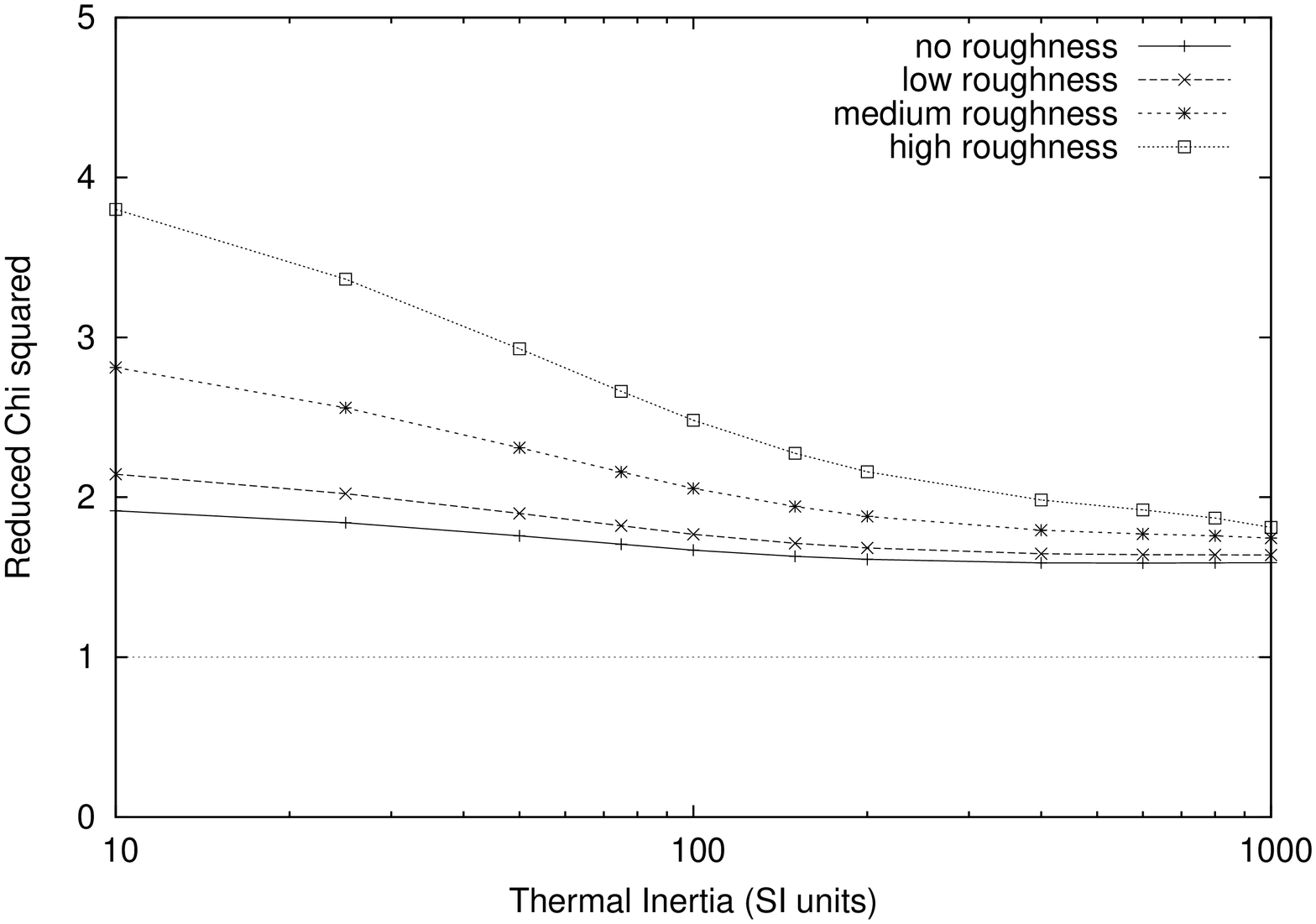}
\end{center}
\caption{As of Fig. \ref{F21} but for the asteroid (720) Bohlinia. \MMP.}
\label{F720}
\end{figure}
}


\begin{thebibliography}{}

\bibitem[Bottke et al.(2005)]{Bottke05}
Bottke,W.F., Durda, D.D., Nesvorn\'y, D., Jedicke, R., Morbidelli, A., Vokrouhlick\'y,
D., Levison, H.F., 2005. Linking the collisional history of the main
asteroid belt to its dynamical excitation and depletion. Icarus 179, 63--94.

\bibitem[Bowell et al.(1989)]{Bowell89}
Bowell, E., Hapke, B., Domingue, D., Lumme, K., Peltoniemi, J., Harris, A.W. 1989.
Application of photometric models to asteroids. In: Binzel, R.P., Gehrels, T., Matthews,
M.S. (Eds.) Asteroids II. Univ. of Arizona Press, Tucson, pp. 524--556.

\bibitem[Bottke et al.(2006)]{Bottke06}
Bottke, W.F., Vokrouhlick\'y, D., Rubincam, D.P., Nesvorn\'y, D., 2006. The
Yarkovsky and YORP effects: Implications for asteroid dynamics. Annu.
Rev. Earth Planet. Sci. 34, 157--191.

\bibitem[Carvano et al.(2007)]{Carvano08}
Carvano, J. M., Barucci, M. A., Fornasier, S., Delbo' M., Lowry, S., and Fitzsimmons, A. 2008.
Surface properties of Rosetta's targets (21) Lutetia and (2867) Steins from ESO observations. Astronomy \& Astrophysics, in press.

\bibitem[Christensen et al.(2003)]{Chris03}
Christensen, P.R., and 21 colleagues. 2003. Morphology and
composition of the surface of Mars: Mars Odyssey THEMIS
results. Science, 300, 2056.

\bibitem[Delbo' (2004)]{Delbo04}
Delbo' M.,
The nature of near-earth asteroids from the study of their thermal infrared emission.
2004.
Freie Universitaet Berlin,
Digitale Dissertation,
on-line at: http://www.diss.fu-berlin.de/2004/289/indexe.html

\bibitem[Delbo' et al.(2007)]{Delbo07}
Delbo', M., Dell'Oro, A., Harris, A.~W., Mottola, S., Mueller, M.\ 2007.\ Thermal inertia of near-Earth asteroids and implications for the magnitude of the Yarkovsky effect.\ Icarus 190, 236--249.

\bibitem[Delbo' and Harris(2002)]{Delbo02}
Delbo', M., Harris, A.~W.\ 2002.\ Physical properties of near-Earth asteroids from
thermal infrared observations and thermal modeling.\ Meteoritics and
Planetary Science 37, 1929--1936.

\bibitem[Durech et al.(2005)]{Dur05}
Durech, J., Grav, T., Jedicke, R., Denneau, L., Kaasalainen, M. 2005. 
Asteroid Models from the Pan-STARRS Photometry. Earth, Moon, and
Planets 97, 179--187.

\bibitem[Hapke (1984)]{Hapke84}
Hapke, B., 1984. Bidirectional reflectance spectroscopy. 3. Correction for
macroscopic roughness. Icarus 59, 41--59.

\bibitem[Harris et al.(2007)]{2007Icar..188..414H} Harris, A.~W., Mueller,
M., Delbo', M., Bus, S.~J.\ 2007.\ Physical characterization of the
potentially hazardous high-albedo Asteroid (33342) 1998 WT$_{24}$ from
thermal-infrared observations.\ Icarus 188, 414--424.

\bibitem[Harris et al.(2005)]{2005Icar..179...95H} Harris, A.~W., Mueller,
M., Delbo', M., Bus, S.~J.\ 2005.\ The surface properties of small
asteroids: Peculiar Betulia - A case study.\ Icarus 179, 95--108.

\bibitem[Jakosky(1986)]{1986Icar...66..117J} Jakosky, B.~M.\ 1986.\ On the
thermal properties of Martian fines.\ Icarus 66, 117--124.

\bibitem[Kaasalainen et al.(2002)]{2002aste.conf..139K} Kaasalainen, M., 
Mottola, S., Fulchignoni, M.\ 2002.\ Asteroid Models from Disk-integrated 
Data.\ Asteroids III 139-150. 

\bibitem[Kaasalainen et al.(2001)]{2001Icar..153...37K} Kaasalainen, M., 
Torppa, J., Muinonen, K.\ 2001.\ Optimization Methods for Asteroid 
Lightcurve Inversion. II. The Complete Inverse Problem.\ Icarus 153, 37-51. 

\bibitem[Kaasalainen and Torppa(2001)]{2001Icar..153...24K} Kaasalainen, 
M., Torppa, J.\ 2001.\ Optimization Methods for Asteroid Lightcurve 
Inversion. I. Shape Determination.\ Icarus 153, 24-36. 

\bibitem[Lebofsky and Spencer (1989)]{Lebo89}
Lebofsky, L.A., Spencer, J.R., 1989. Radiometry and thermal modeling of asteroids.
In: Binzel, R.P., Gehrels, T., Matthews, M.S. (Eds.), Asteroids II.
Univ. of Arizona Press, Tucson, pp. 128--147.

\bibitem[Lebofsky et al.(1986)]{1986Icar...68..239L} Lebofsky, L.~A.,
Sykes, M.~V., Tedesco, E.~F., Veeder, G.~J., Matson, D.~L., Brown, R.~H.,
Gradie, J.~C., Feierberg, M.~A., Rudy, R.~J.\ 1986.\ A refined 'standard'
thermal model for asteroids based on observations of 1 Ceres and 2 Pallas.\
Icarus 68, 239--251.

\bibitem[Mellon et al. (2000)]{Mellon00}
Mellon, M.T., Jakosky, B.M., Kieffer, H.H., Christensen, P.R. 2000. High-resolution thermal
inertia mapping from the Mars global surveyor thermal emission spectrometer.
Icarus 148, 437--455.

\bibitem[Mignard et al. (2007)]{MignardGaiaGLO07}
Mignard, F., Cellino, A., Muinonen, K., Tanga, P., Delbo', M., Dell'Oro,
A., Granvik, M., Hestroffer, D., Mouret, S., Thuillot, W., Virtanen, J., 2008.
The Gaia mission: Expected applications to asteroid science. Earth Moon
Planets, 101, 97--125.

\bibitem[Michikami et al.(2007)]{2007P&SS...55...70M} Michikami, T.,
Moriguchi, K., Hasegawa, S., Fujiwara, A.\ 2007.\ Ejecta velocity
distribution for impact cratering experiments on porous and low strength
targets.\ Planetary and Space Science 55, 70--88.

\bibitem[Morbidelli and Vokrouhlick\'y (2003)]{Morby03}
Morbidelli, A., Vokrouhlick\'y, D., 2003. The Yarkovsky-driven origin of near-
Earth asteroids. Icarus 163, 120--134.

\bibitem[Mueller (2007)]{Migo07}
Mueller, M. Surface Properties of Asteroids from Mid-Infrared Observations and Thermophysical Modeling. 2007. Freie Universitaet Berlin, Digitale Dissertation, on-line at:
http://www.diss.fu-berlin.de/2007/471/indexe.html

\bibitem[Mueller et al. (2006)]{Migo06}
Mueller, M., Harris, A.W., Bus, S.J., Hora, J.L., Kassis, M., Adams, J.D. 2006. The
size and albedo of Rosetta fly-by target 21 Lutetia from new IRTF measurements and
thermal modeling. Astronomy \& Astrophysics 447, 1153--1158.

\bibitem[Mueller et al. (2004)]{Migo04}
Mueller, M., Delbo', M., di Martino, M., Harris, A.W., Kaasalainen, M., Bus, S.J. submitted
in 2004. Indications for regolith on Itokawa from thermal-infrared observations.
ASP Conference Series, in press.

\bibitem[M{\"u}ller et al.(2005)]{2005A&A...443..347M} M{\"u}ller, T.~G., Sekiguchi, T., Kaasalainen, M., Abe, M., Hasegawa, S.\ 2005.\ Thermal infrared observations of the Hayabusa spacecraft target asteroid (25143) Itokawa.\ Astronomy and Astrophysics 443, 347--355. 

\bibitem[M{\"u}ller et al.(2004)]{2004A&A...424.1075M} M{\"u}ller, T.~G.,
Sterzik, M.~F., Sch{\"u}tz, O., Pravec, P., Siebenmorgen, R.\ 2004.\
Thermal infrared observations of near-Earth asteroid 2002 NY40.\ Astronomy
and Astrophysics 424, 1075--1080.

\bibitem[M{\"u}ller and Blommaert(2004)]{2004A&A...418..347M} M{\"u}ller,
T.~G., Blommaert, J.~A.~D.~L.\ 2004.\ 65 Cybele in the thermal infrared:
Multiple observations and thermophysical analysis.\ Astronomy and
Astrophysics 418, 347--356.

\bibitem[M\"uller and Lagerros(1998)]{1998A&A...338..340M} M\"uller, T.~G.,
Lagerros, J.~S.~V.\ 1998.\ Asteroids as far-infrared photometric standards
for ISOPHOT.\ Astronomy and Astrophysics 338, 340--352.

\bibitem[Spencer et al.(1989)]{Spencer89} Spencer, J.~R.,
Lebofsky, L.~A., Sykes, M.~V.\ 1989.\ Systematic biases in radiometric
diameter determinations.\ Icarus 78, 337--354.

\bibitem[Tedesco et al. (2004)]{Tedesco04}
Tedesco, E.F., P.V. Noah, M. Noah, and S.D. Price. 2004. IRAS Minor Planet Survey. IRAS-A-FPA-3-RDR-IMPS-V6.0. NASA Planetary Data System.
http://www.psi.edu/pds/resource/imps.html

\bibitem[Tedesco et al. (2002)]{Tedesco02}
Tedesco, E.F., Noah, P.V., Noah, M., Price, S.D. 2002. The Supplemental IRAS Minor
Planet Survey. Astron. Journal 123, 1056--1085.

\bibitem[Tedesco (1992)]{Tedesco92}
Tedesco, E.F. (Ed.) 1992. IRAS Minor Planet Survey. Phillips Laboratory Technical Report
No. PL--TR--92--2049. Hanscom Air Force Base, Massachusetts.

\bibitem[Trilling et al.(2007)]{2007DPS....39.3515T} Trilling, D.~E.,
Bhattacharya, B., Blaylock, M., Stansberry, J.~A., Sykes, M.~V., Wasserman,
L.~H.\ 2007.\ The Spitzer Asteroid Catalog: Albedos And Diameters of 35,000
Asteroids.\ AAS/Division for Planetary Sciences Meeting Abstracts 39,
\#35.15.

\bibitem[Walsh and Richardson(2006)]{2006Icar..180..201W} Walsh, K.~J., 
Richardson, D.~C.\ 2006.\ Binary near-Earth asteroid formation: Rubble pile 
model of tidal disruptions.\ Icarus 180, 201--216. 

\bibitem[Vokrouhlick{\'y} et al.(2003)]{2003Natur.425..147V} 
Vokrouhlick{\'y}, D., Nesvorn{\'y}, D., Bottke, W.~F.\ 2003.\ The vector 
alignments of asteroid spins by thermal torques.\ Nature 425, 147--151. 


\end{thebibliography}
\end{document}